\documentclass[fleqn,11pt,twoside]{article}

\usepackage{amsthm,amsthm,amssymb, color, xcolor,epsfig, graphics, subfigure}

\usepackage{amsmath, graphicx, latexsym, lscape}

\usepackage{cite}

\usepackage{changes}
\usepackage{tabularx}
\usepackage{diagbox}
\usepackage{makecell}
\usepackage{xcolor}
\usepackage{soul}
\RequirePackage{graphicx}
\RequirePackage{flushend}
\usepackage{bbm}
\usepackage{bm}
\usepackage{booktabs}  

\makeatletter
\newcommand{\copyrightnote}[2]{{\renewcommand{\thefootnote}{}
 \footnotetext{\small\it
\begin{flushleft}
 \copyright \ #1   #2  
\end{flushleft}}}}

\newcommand{\Name}[1]{\begin{flushleft}
                       \LARGE \bf #1
                       \end{flushleft}\vspace{-3mm}}

\newcommand{\Author}[1]{\begin{flushleft}
                       \it #1 \end{flushleft}}

\newcommand{\Address}[1]{\begin{flushleft}
                       \it #1 \end{flushleft}}

\newcommand{\Date}[1]{\begin{flushleft}
                      \small  \it #1 \end{flushleft}}

\newcommand{\evenhead}{Author \ name}
\newcommand{\oddhead}{Article \ name}

\renewcommand{\@evenhead}{
\hspace*{-3pt}\raisebox{-15pt}[\headheight][0pt]{\vbox{\hbox to \textwidth
{\thepage \hfil \evenhead}\vskip4pt \hrule}}}
\renewcommand{\@oddhead}{
\hspace*{-3pt}\raisebox{-15pt}[\headheight][0pt]{\vbox{\hbox to \textwidth
{\oddhead \hfil \thepage}\vskip4pt\hrule}}}
\renewcommand{\@evenfoot}{}
\renewcommand{\@oddfoot}{}

\newlength{\tagwidth}
\RequirePackage{fix-cm}

\long\def\@makecaption#1#2{%
  \vskip\abovecaptionskip
  \sbox\@tempboxa{\small \textbf{#1.}\ \ #2}%
  \ifdim \wd\@tempboxa >\hsize
    {\small \textbf{#1.}\ \ #2}\par
  \else
    \global \@minipagefalse
    \hb@xt@\hsize{\hfil\box\@tempboxa\hfil}%
  \fi
  \vskip\belowcaptionskip}

\newcommand{\JNMPnumberwithin}[3][\arabic]{%
  \@ifundefined{c@#2}{\@nocounterr{#2}}{%
    \@ifundefined{c@#3}{\@nocnterr{#3}}{%
      \@addtoreset{#2}{#3}%
      \@xp\xdef\csname the#2\endcsname{%
        \@xp\@nx\csname the#3\endcsname .\@nx#1{#2}}}}%
}

\newcommand{\resetfootnoterule} {
  \renewcommand\footnoterule{%
  \kern-3\p@
  \hrule\@width.4\columnwidth
  \kern2.6\p@}
}

\renewcommand{\footnoterule}{}

\makeatother

\theoremstyle{definition}

\begin{document}

\renewcommand{\evenhead}{ {\LARGE\textcolor{blue!10!black!40!green}{{\sf \ \ \ ]ocnmp[}}}\strut\hfill 
A Latifi, M A Manna and R A Kraenkel
}
\renewcommand{\oddhead}{ {\LARGE\textcolor{blue!10!black!40!green}{{\sf ]ocnmp[}}}\ \ \ \ \  
Wind-wave interaction in finite depth
}

%%%% Matter for the first page 
\thispagestyle{empty}
\newcommand{\FistPageHead}[3]{
\begin{flushleft}
\raisebox{8mm}[0pt][0pt]
{\footnotesize \sf
\parbox{150mm}{{Open Communications in Nonlinear Mathematical Physics}\ \ \ \ {\LARGE\textcolor{blue!10!black!40!green}{]ocnmp[}}
\quad Special Issue 2, 2024\ \  pp
#2\hfill {\sc #3}}}\vspace{-13mm}
\end{flushleft}}

\FistPageHead{1}{\pageref{firstpage}--\pageref{lastpage}}{ \ \ }

\strut\hfill

\strut\hfill

\copyrightnote{The author(s). Distributed under a Creative Commons Attribution 4.0 International License}

\begin{center}

{\bf {\large Proceedings of the OCNMP-2024 Conference:\\ 

\smallskip

Bad Ems, 23-29 June 2024}}
\end{center}

\smallskip

\Name{	Wind-wave interaction in finite depth: linear and nonlinear approaches, blow-up and soliton breaking in finite time, integrability perspectives}

\Author{A.~Latifi }

\Address{Department of Mechanics, Faculty of Physics,\\	Qom University of Technology \\Qom, Iran}

\Author{M.A.~ Manna }

\Address{	Universit\'e de Montpellier (UM)\\ and \\Conseil National de la Recherche Scientifique (CNRS),\\
	Laboratoire Charles Coulomb UMR 5221\\
	F-34095, Montpellier, France }

\Author{R.A.~Kraenkel}
\Address{Instituto de F\'isica T\'eorica, UNESP \\
	Universidade Estadual Paulista \\ 
	01140-070, S\~ao Paulo, Brazil}

\Date{Received June 21, 2024; Accepted July 24, 2024}

\setcounter{equation}{0}

\begin{abstract}

\noindent 
	This work is a review of our recent analytical advances of the evolution of surface water solitary waves in Miles and Jeffreys’ theories of wind wave interaction in water of finite depth. 
Although many works have been conducted based on Miles and Jeffreys’ approach, only a few studies have been carried out on finite depth. 
The present review is divided into two major parts. The first corresponds to the surface water waves in a linear regime and its nonlinear extensions. In this part, Miles’ theory of wave amplification by wind is extended to the case of finite depth. The dispersion relation provides a wave growth rate depending on depth. A dimensionless water depth parameter, depending on the depth and a characteristic wind speed, induces a family of curves representing the wave growth as a function of the wave phase velocity and the wind speed. Our theoretical results are in good agreement with the data from the Australian Shallow Water Experiment and the data from the Lake George experiment. 
In the second part of this study, Jeffreys’ theory of wave amplification by wind is extended to the case of finite depth, where the fully nonlinear focusing Serre-Green-Naghdi (SGN) equation is derived. "Anti-dissipation" occurs due to the continuous transfer of wind energy to water surface waves. We find the solitary wave solution of the system, with an increasing amplitude under the action of the wind. This continuous increase in amplitude leads to the "soliton" breaking and blow-up of the surface wave in finite time for infinitely large asymptotic space. This dispersive, focusing and fully nonlinear phenomenon is equivalent to the linear instability at infinite time. The theoretical blow-up time is calculated based on actual experimental data.
By applying an appropriate perturbation method, the SGN equation yields a focusing weakly nonlinear dispersive Korteweg–de Vries–Burger-type  (KdV-B) equation. We show that the continuous transfer of energy from wind to water results in the growth of the KdV-B soliton-like’s amplitude, velocity, acceleration, and energy over time while its effective wavelength decreases. This phenomenon differs from the classical results of Jeffreys’ approach due to finite depth. Again, blow-up and breaking occur in finite time. These times are calculated and expressed for solitary wave solution- and wind-appropriate parameters and values. These values are measurable in usual experimental facilities. The kinematics of the breaking is studied, and a detailed analysis of the breaking kinetics and  breaking time is conducted using various criteria. Finally, some integrability perspectives are presented.

\end{abstract}

\label{firstpage}

%%%% The Article text starts here

\section{Introduction}

The Navier-Stokes equation for air and water is the starting point of wind-generated water waves. Since the Navier-Stokes equation is analytically unsolved,  various approximations and assumptions are required to obtain particular solutions.

The wind wave growth is due to the continuous energy and momentum flow from the air to the water. Winds generate surface waves; in turn, surface water waves modify the airflow. In this way, the atmosphere depends on the wave state near the water surface. The physical mechanism behind is "focusing" with a continuous energy and momentum flow from the air to the surface wave generating an exponential growth of the wave amplitude
	in time, more or less quickly, depending on the wind speed and the water depth.

However, waves may lose energy because of dissipation. The {\it action balance equation}, a fundamental equation in fluid dynamics, commonly represents this temporal and spatial dynamics. In deep water, it reads  \cite{Janssen}:
\begin{equation}
	\frac{\partial }{\partial t}\mathcal{N} +   \overrightarrow{\nabla}.(\overrightarrow{c_g}\mathcal{N})= S, %= S_{in}+S_{nl}+S_{ds},
\end{equation}\label{source}
where $\overrightarrow{c_g}$ is the group velocity observed in a frame moving with the wave, $\mathcal{N}$ is {\it the action density} defined as follows:
\begin{equation}
	\mathcal{N}=\frac{\mathcal{E}}{\omega},
\end{equation}
where $\omega$ is the wave frequency, $ \mathcal{E} $ is the energy density

\begin{equation}
	\mathcal{E}=2\rho_w g \eta_0^2,
\end{equation}
where $\eta_0$ the small wave amplitude, and $S$ is the source term
\begin{equation}\label{source1}
	S=S_{in}+S_{nl}+S_{ds}+\ldots,
\end{equation} 
where $S_{in}$, $S_{nl}$, $S_{ds}$ are the effects of the wind input, non-linear interactions and dissipation due to white capping, respectively.

The pioneer theoretical works on surface wind-wave growth were essentially done by Jeffreys and Miles \cite{Jeffreys1, Jeffreys2, Phillips, Miles2}, till some more recent studies\cite{Janssen2,Belcher}. These works essentially aim to compute the term
$S_{in}$ in Eq. \eqref{source1}. Later, numerical approaches attempted to campute $S_{in}$, $S_{nl}$, $S_{ds}$  \cite{Janssen}.

However,  all theories mentioned above are focused on {\it deep water domain} and  are not well adapted to 
correctly describe wind-waves
generation near-shore oceans or shallow lakes. 
This limitation challenges both the physics and engineering
communities. Indeed, in  the {\it finite depth water domain} the source term is
\begin{equation}
	S=S_{in}+S_{nl}+S_{ds}+S_{bf}+S_{tri}+ \ldots.
\end{equation}
where $S_{bf}$ is the bottom friction,  and $S_{tri}$ is triad nonlinear wave interactions. Moreover, 
$S_{in}$ being strongly influenced by the finite depth $h$, must be recalculated.
To our knowledge, although many works have been conducted based on Miles and Jeffreys’ approach, only a few studies are carried out on finite depth \cite{JAFD_2017, Manna_2018, fluids7080266, fluids8080231}.

This review is the result of our analytical extension of the evolution of surface water solitary waves in Miles and Jeffreys’ theories of wind wave interaction in water of finite depth with the Euler equations as outset. The analytic approach is essential for further numerical investigations due to the scale of energy dissipation near coasts. Indeed, the scale of energy dissipation is of the order of a micrometre, which requires $  10^{25} $ mesh nodes to produce correct predictions on scales of 100 km. Hence, no pure numerical modelling of this problem without recourse to theoretical developments has a chance of succeeding.

Based on the  our latest progress \cite{JAFD_2017,Manna_2018,fluids7080266,fluids8080231}, this work aims to give {\it l'etat de l'art } in the field of wind-wave interactions in finite depth, and thereby supply  a theoretical basis allowing to go beyond the empirical laws. A permanent concern in this work is to compare our theoretical results to experimental data or check the feasibility of experimentation and field observations.

This paper is divided into two parts: The first part is dedicated to the extension of Miles' approach \cite{Miles2} to finite depth, where the linearized Euler equations in the water domain are coupled with weak nonlinearity to linearized Euler equations in the air domain. The problem is solved at the interface, and the linear dispersion relation of wave amplification at finite depth is calculated. By introducing dimensionless variables and scaling, the wind wave's growth rate is obtained. Our theoretical laws are compared with both the Young-Verhagen data and plots of empirical relationships from the Lake George experiment and with Donelan's data from the AUSWEX program \cite{YoungVerhagen1,YoungVerhagen2,IjimaTang}.

Introducing a simple {\it modus operandi} for the Miles' mechanism allows us to derive an anti-diffusive nonlinear Schr\"odinger equation for the wind-wave in finite depth and we derive the Akhmediev, Peregrine, and Ma solutions for weak wind inputs.

The second part of this paper is devoted to the extension of Jeffrey's mechanism \cite{Jeffreys1,Jeffreys2} where in different scales,  the focusing KdV-B equation and the fully nonlinear SGN are derived. The focusing KdV-B equation is obtained by the coupling of the weakly nonlinear Euler equations in the water domain to the linearized Euler equations in the air domain. In contrast, the fully nonlinear SGN is derived by coupling the nonlinear Euler equations in the water domain to the linearized Euler equations in the air domain.

For both SGN and KdV-B equations, solitary waves solutions exhibiting blow-up in finite time are found. By considering its kinematics description, the solition breaking phenomenon is analyzed in detail in the case of KdV-B equation.

%%%%%%%%%%%%%%%%%%%%%%%%%%%%%%%%%%%%%%%%%%%%%%%%%%%%%%%%%%%%%%%%%%%%%%%%%%%%%%%%%%%%%%%%%%%%%%%%%%%%%%%%%%%%%%%%%%%%%%%%

This review is organized as follows: Section 2 studies the mutual action of linearised water and air dynamics through Mile's mechanism in finite depth. The wave growth rate is analytically calculated and compared to field experiment results and empiric laws. A good agreement between theoretical and experimental results is observed. Section 3 compares our theoretical results to experimental and empirical laws. In Section 3, we apply the Miles approach from a quasi-linear perspective. Here, the air dynamics remain linear, while the water domain is viewed as irrotational and nonlinear. This modification gives rise to the Nonlinear Schr\"odinger equation, from which we derive the Akhmedeiev, Peregrine, and Ma solutions solutions for weak wind input. Sections 4 and 5 are devoted to the nonlinear Jeffrey's approach. Namely, in section 3, the water domain is considered irrotational and nonlinear, and the air domain is linearized. Jeffrey's sheltering mechanism allows us to derive the SGN equation in finite depth, which we solve analytically. The solution is a solitary wave with the remarkable feature of blowing up in finite time. The blow-up time is calculated based on experimental measurement. In section 5, applying an appropriate approximation, we derive the KdV-B equation from the SGN equation we solve analytically. Again, the solution is a solitary wave with a blow-up in finite time. In this section, we conduct a detailed analysis of the wave breaking before the blow-up, which we compare to existing breaking criteria. In Section 6, we present some integrability perspectives. Section 7 summarises our results, draws conclusions, and gives perspectives on further developments. In the Appendix, the direct derivation of the KdV-B equation Euler equation is shown.

\section{Mile's Mechanism}

\begin{figure}
	\centering
	\includegraphics[width=0.7\linewidth]{"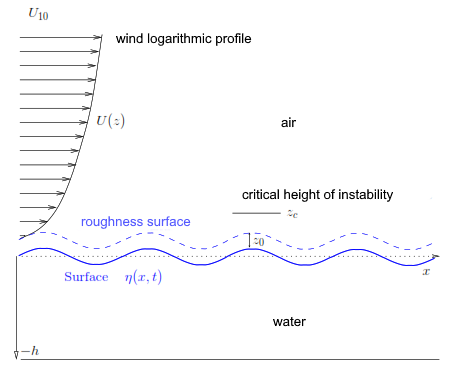"}
	\caption{The Miles mechanism is based on a
		particular interaction between air flows and surface waves. It's about a linear mechanism of resonance between the wind speed and the water phase speed, which will induce a change in pressure, which in turn will modify the propagation properties of linear water surface waves. We consider two fluids, air and water, separated by a free surface $ (x, t) $. The bottom is located at a depth $ - h $, and we prescribe a wind $  U(z) $ having a specific profile, typically logarithmic. The interface at rest. The perturbed air-water interface will be described by $z=\eta(x,t)$.}
	\label{fig:mile-figure}
\end{figure}

The Miles mechanism in infinite or finite depth is based on a
particular interaction between air flows and surface waves. It's about a linear mechanism of resonance between the wind speed and the water phase speed, which will induce a change in pressure and modify the propagation properties of linear water surface waves. We therefore model this instability as follows: either a coordinate system Cartesian dimensions with $ x $ the horizontal, $ z $ the height and $ y $ the transverse dimension, which we will not consider here. We consider two fluids, air and water, separated by a free surface $ (x, t) $. The constant densities of air and water are $ \rho_a $ and $ \rho_w $. Pressure fields in water and air are denoted respectively $ P_w $ and $ P_a $. The bottom is located at a depth  $z=-h$ and the interface $z=\eta(x,t)$. We prescribe a wind $  U(z) $ having a specific profile, typically logarithmic, as in Figure \eqref{fig:mile-figure}. Therefore, we will write Euler's equations for air and water, with an additional condition of pressure continuity connecting them.  We assume the dynamic to be linear and disregard the air turbulence, building a \textit{quasi-laminar theory}.

\subsection{The linearized water dynamics}
\label{water domain}

In the water domain, we consider the Euler equations for finite depth.
The horizontal and vertical velocities of the fluid are $u(x,z,t)$ and $w(x,z,t)$.
The continuity equation and the linearized equation of motion in the water domain read  \cite{Lighthill1978-fo}:
\begin{equation}\label{EulerWater-1}
	u_x + w_z = 0, \quad \rho_w u_t= -P_{x},\quad \rho_w w_t=-P_{z}-g \rho_w,
\end{equation}
where $P(x,z,t)$ is the pressure, $g$ the gravitational acceleration, $\rho_w$ is the water density and
subscripts in $u$, $w$, $ P_x $ and $P_z$ denote partial derivatives of $ P $ with respect to $ x $ and $ z $, receptively.
The boundary conditions at $z = -h $ and at $z = \eta (x,t)$ are
\begin{eqnarray}
	&&w(-h)= 0,\quad \eta_{t}=w(0), \label{EulerWaterBC1}\\
	&&	P(x,\eta,t)= P_a(x,\eta,t),\label{EulerWaterBC2-1}
\end{eqnarray}
where $P_a$ is the air pressure evaluated at $z=\eta$. Thus equation (\ref{EulerWaterBC2-1}) is 
the continuity of the pressure across the air/water interface. As this is a vital assumption for the growth mechanism, we give it a 
more pleasant expression. So, let us introduce a reduced pressure defined by
\begin{equation}\label{reduced pressure}
	\mathbf{P}(x,z,t)=P(x,z,t) + \rho_w gz -P_{0},
\end{equation}
where $P_0$ is the atmospheric pressure. 
In terms of (\ref{reduced pressure}) 
equations (\ref{EulerWater-1})-(\ref{EulerWaterBC2-1}) read
\begin{equation}\label{EulerWaterReduced}
	u_x + w_z = 0, \quad \rho_w u_t= -\mathbf{P}_x,\quad \rho_w w_t=-\mathbf{P}_z,
\end{equation}
\begin{eqnarray}
	&&	w(-h)= 0,\quad \eta_{t} = w(0), \label{EulerWaterReducedBC1} \\
	&&	\mathbf{P}(x,\eta,t)=P_a(x,\eta,t)+\rho_w g\eta - P_0. \label{EulerWaterReducedBC2}
\end{eqnarray}
The linear equations system (\ref{EulerWaterReduced})-(\ref{EulerWaterReducedBC2}) can be solved, assuming normal mode solutions 
as
\begin{eqnarray}\label{Fourier}
	\mathbf{P}&=&\mathcal{P}(z)\exp{(i\theta)},\quad u=\mathcal{U}(z)\exp{(i\theta)},\nonumber\\
	\quad w&=&\mathcal{W}(z)\exp{(i\theta)},\quad \eta=\eta_{0}\exp{(i\theta)},
\end{eqnarray}
with $\theta= k(x-ct)$ where $k$ is the wavenumber, $c$ the phase speed and $\eta_0$ is a constant.
Using equations (\ref{EulerWaterReduced}), (\ref{EulerWaterReducedBC1}), 
(\ref{EulerWaterReducedBC2}) and (\ref{Fourier}) we obtain 
\begin{eqnarray}
	w(x,z,t) &=&\frac{-ikc\sinh{k(z+h)}}{\sinh{kh}}\eta_0\exp{(i\theta)},\label{wlinear}\\
	u(x,z,t) &=& \frac{kc\cosh{k(z+h)}}{\sinh{kh}}\eta_0\exp{(i\theta)}, \label{ulinear}\\
	\mathbf{P}(x,z,t) &=& \frac{k\rho_wc^2\cosh{k(z+h)}}{\sinh{kh}}\eta_0\exp{(i\theta)}\label{Plinear}.
\end{eqnarray}

The phase speed $c$ is unknown in equations (\ref{wlinear})-(\ref{Plinear}). 
To determine $c$ we have to consider the boundary conditions (\ref{EulerWaterReducedBC2}), not yet used, and (\ref{EulerWaterReducedBC1}) which yields
\begin{equation}\label{deterc}
	\rho_w\eta_0\exp{(i\theta)}\{c^2k\coth{kh}- g\} + P_0=P_a(x,\eta,t).
\end{equation}

In the single-domain problem $P_a(x,\eta,t)= P_0$ and (\ref{deterc}) gives the usual expression for c,
\begin{equation}
	c^2 = c_0^2 =\frac{g}{k}\tanh{(kh)}.
\end{equation}
It is not the case in the problem under consideration in the present paper in which the determination of c needs the use of 
the air pressure evaluated at  $z=\eta$. 
%%%%%%%%%%%%%%%%%%%%%%%%%%%%%%%%%%%%%%%%%%%%%%%%%%%%%%%%%%%%%%%%%%%%%%%%%%%%%%%%
\subsection{The linearized air dynamics}
\label{air domain}
Let us consider the linearized governing equation of a steady airflow, with a prescribed mean horizontal velocity $U(z)$ depending on the 
vertical coordinate $z$. We are going to study perturbations to the mean 
flow $U(z)$: $u_a(x,z,t)$, $w_a(x,z,t)$ and $P_a(x,z,t)$ where subscript $a$ stands for \emph{air}. So with 
$\mathbf{P}_a(x,z,t)=P_a(x,z,t) + \rho_a gz -P_{0} $, $\rho_a$ the air 
density, and $U'= dU(z)/dz$ we have the following equations
\begin{eqnarray}
	&&u_{a,x}+w_{a,z}=0,\label{continuityair}\\
	&&	\rho_a[u_{a,t} + U(z)u_{a,x} + U'(z)w_a]=-\mathbf{P}_{a,x},\label{uair}\\
	&&	\rho_a[w_{a,t} + U(z)w_{a,x}]=-\mathbf{P}_{a,z},\label{wair}
\end{eqnarray}
which must be completed with the appropriate boundary conditions. The first is the kinematic boundary condition for air,
evaluated at the aerodynamic sea surface roughness $z_0$ located just above the interface. Through this paper, $z_0$ will be 
regarded as a constant independent from the sea state. This is a widely used approximation, first proposed by 
\cite{Charnock}. For the datasets used later on, the wind speed ranges are such that the roughness may be seen as a 
constant \cite{Fairall}. The kinematic boundary condition reads
\begin{equation}\label{etaair}
	{\eta}_t + U(z_0){\eta}_x = w_a(z_0).
\end{equation}
We choose $U(z)$ as the logarithmic wind profile. This is commonly used to describe the vertical distribution of the horizontal mean 
wind speed within the lowest portion of the air side of the marine boundary 
layer \cite{Garratt}. It can also be justified with scaling arguments and solution matching between the near-surface and geostrophic air layer \cite{Tennekes}.
\begin{equation}\label{Udefiniton}
	U(z) = U_{1} \ln(z/z_{0}),\quad U_{1} = \frac{u_{*}}{\kappa},\quad \kappa 
	\approx 0.41,
\end{equation}
where $u_{*}$ is the friction velocity  
and $\kappa$ the Von K\'{a}rm\'{a}n constant. So, eq. (\ref{etaair}) can be reduced to 
\begin{equation}\label{etaairreduced}
	\eta_t = w_a(z_0).
\end{equation}
This equation describes the influence of the surface perturbation on the 
vertical perturbed wind speed. Next
we assume $\mathbf{P}_a=\mathcal{P}_a(z)\exp{(i\theta)}$, $u_a=\mathcal{U}_a(z)
\exp{(i\theta)}$, $w_a=\mathcal{W}_a(z)\exp{(i\theta)}$ and we add the following 
boundary conditions on $\mathcal{W}_a$ and $\mathcal{P}_a $, 
\begin{eqnarray}
	&&	\lim\limits_{ z \to +\infty}(\mathcal{W}_a' + k\mathcal{W}_a) = 0, \label{W_ainfinite}\\
	&&	\lim\limits_{ z \to z_0} \mathcal{W}_a = W_{0},\label{W_ainz0}\\
	&&	\lim\limits_{ z \to+\infty} \mathcal{P}_a = 0, \label{P_ainfinite}
\end{eqnarray}
that is, the disturbance plus its derivative vanish at infinity, and the vertical component of 
the wind speed is enforced by the wave movement at the sea surface.
Then, using
equations (\ref{continuityair})-(\ref{wair}) and (\ref{P_ainfinite}) we obtain
\begin{eqnarray}
	w_a(x,z,t)&=&\mathcal{W}_a\exp{(i\theta)},\label{w_a}\\
	u_a(x,z,t)&=&\frac{i}{k}\mathcal{W}_{a,z}\exp{(i\theta)},\label{u_a}\\
	\mathbf{P}_a(x,z,t)&=&ik\rho_a\exp{(i\theta)}\int^{\infty}_{z}[U-c]
	\mathcal{W}_{a}dz'.\label{mathP_a}
\end{eqnarray}
Removing the pressure from the Euler equations, we find the well-known Rayleigh equation  \cite{Rayleigh}
$\forall z~ \backslash~ z_{0} < z < +\infty$ (inviscid Orr-Sommerfeld equation)
\begin{equation}\label{Rayleigh}
	(U - c) (\mathcal{W}_a'' - k^2 \mathcal{W}_a) - U'' \mathcal{W}_a = 0
\end{equation}
which is singular at the critical, or matched height $z_{c} = z_0 e^{c \kappa/ u_{*}} > z_{0} > 0$, where 
$U(z_{c}) = c$. We recall that this model disregards any kind of turbulence, and so the critical height is 
set above any turbulent eddies or other non-linear phenomena. In equations (\ref{w_a})-(\ref{Rayleigh}) neither $\mathcal{W}_a(z)$ nor $c$ are known. 
In order to find $c$, we have to calculate
$P_a(x,\eta,t)$. We obtain
\begin{eqnarray}\label{integralforP}
	&&	P_a(x,\eta,t)=P_0 - \rho_a g \eta 
	+ {ik\rho_a\exp{(i\theta)}\int^{\infty}_{z_0}[U(z)-c]\mathcal{W}_a(z)dz},
\end{eqnarray}
where the lower integration bound is taken at the constant roughness height $z_0$ 
instead of $z=\eta$ since we are studying the linear problem.
Using
equation (\ref{etaairreduced}) to eliminate the term $ik\rho_a\exp(i\theta)$
the equation (\ref{integralforP}) in (\ref{deterc}) yields
\begin{equation}\label{equationforc1}
	g(1-s) + c\frac{sk^2}{W_0}I_1-c^2\{\frac{sk^2}{W_0}I_2 + k\coth(kh)\}=0,
\end{equation}
where $s=\rho_a/\rho_w$ and the integrals $I_1 $ and $I_2$ are defined as follow
\begin{equation}
	I_1=\int_{z_0}^{\infty} U\mathcal{W}_a dz,\quad I_2=\int_{z_0}^{\infty}\mathcal{W}_a dz.
\end{equation}
Equation (\ref{equationforc1}) is the dispersion relation of the problem. The parameter $s$ is small ($\rho_a/\rho_w \sim 10^{-3}$) and
(\ref{equationforc1}) may be approximated as
\begin{equation}
	c=c_0 + sc_1 + O(s^2).
\end{equation}
The explicit form of $c_1$ is calculated in the next section. 
Therefore, we can find $\mathcal{W}_a(z)$ by solving (\ref{Rayleigh}) with $c$ replaced by $c_0$, that is to say, of order 
zero in $s$. 
\subsection{The wave growth rate}
\label{growth}
The function $\mathcal{W}_a(z)$ 
is complex and consequently $c$ also. Its imaginary part gives the growth rate of $\eta(x,t)$ defined by
\begin{equation}
	\gamma =k\Im{(c)},
\end{equation}
where $\Im{(c)}$ is the imaginary part of $c$.
The theoretical and numerical results concerning the growth rate $\gamma$ are studied and computed with two dimensionless parameters $\delta$ (see \cite{YoungVerhagen1} 
and \cite{YoungVerhagen2}) and $\theta_{dw}$ defined by
\begin{equation}\label{delta}
	\delta = \frac{gh}{U^2_1},\quad \theta_{dw}=\frac{1}{U_1}\sqrt{\frac{g}{k}}.
\end{equation}
The dimensionless parameter $\delta$, for constant $ U_1$, measures the influence of the finite fluid depth on the rate of growth of
$\eta(x,t)$. The parameter $\theta_{dw}$ can be seen as a {\it theoretical analogous of the deep water wave age}.
It measures the relative value of the deep water phase speed about the characteristic wind velocity $U_1$.
Now a {\it theoretical analogous of the  finite depth wave age} $\theta_{fd}$ can be introduced as
\begin{equation}\label{waveagefinite}
	\theta_{fd}=\frac{1}{U_1}\sqrt{\frac{g}{k}}\sqrt{\tanh({kh})}=
	\theta_{dw} T^{1/2},
\end{equation}
where 
\begin{equation}\label{tanh}
	T = \tanh(\frac{\delta}{\theta_{dw}^2}).
\end{equation} 
The form (\ref{waveagefinite}) is a depth weighted parameter such that for a finite and constant
$\theta_{dw}$ we have $\theta_{fd}\sim \theta_{dw}$ if $\delta \rightarrow \infty$ and
$\theta_{fd}\sim \delta^{1/2}=\sqrt{gh}/U_1$ if $\delta \rightarrow 0$. 
To obtain the growth rate, we introduce
the following non-dimensional variables and scalings, hats meaning dimensionless quantities
\begin{eqnarray}\label{parameters}
	U&=&U_1\hat{U},\quad \mathcal{W}_a=W_0\hat{\mathcal{W}_a},
	\quad z=\frac{\hat{z}}{k}, \nonumber\\
	&& c = U_{1} \hat{c},\quad t = \frac{U_{1}}{g} \hat{t}.
\end{eqnarray}
Using (\ref{delta}) and (\ref{parameters}) in equation (\ref{equationforc1}) and retaining only the terms of order one in $s$ we obtain $c$,
\begin{eqnarray}
	\hat{c}=\hat{c}(\delta,\theta_{dw})=\theta_{dw}T^{1/2} -\frac{s}{2} \theta_{dw}T^{1/2} 
	+\frac{s}{2}\{ T\hat{I}_1 -
	\theta_{dw} T^{3/2} \hat{I}_2 \},
\end{eqnarray}
and with $e^{\gamma t} = e^{k {\Im}(c) t}= e^{\Im{(\hat{c})} \hat{t}/\theta_{dw}^{2}}$, we have the dimensionless growth 
rate $\hat{\gamma} = \frac{U_{1}}{g}\gamma$ as,
\begin{equation}
	\hat{\gamma}=\frac{s}{2}\{ \frac{T \Im{(I_1)}}{\theta^2_{dw}}-
	\frac{T^{3/2}\Im{(I_2)}}{\theta_{dw}}\},
	\label{growthrate}
\end{equation}
So, we can compute it for a given $(\delta,\theta_{dw})$ set.
The $\delta$ parameter does not appear explicitly, allowing us to compute $\gamma$ for an infinite depth, where we have just $T\rightarrow 1$. This gives back to Miles' theory.

\begin{figure}
	\centering
	\includegraphics[width=0.7\linewidth] {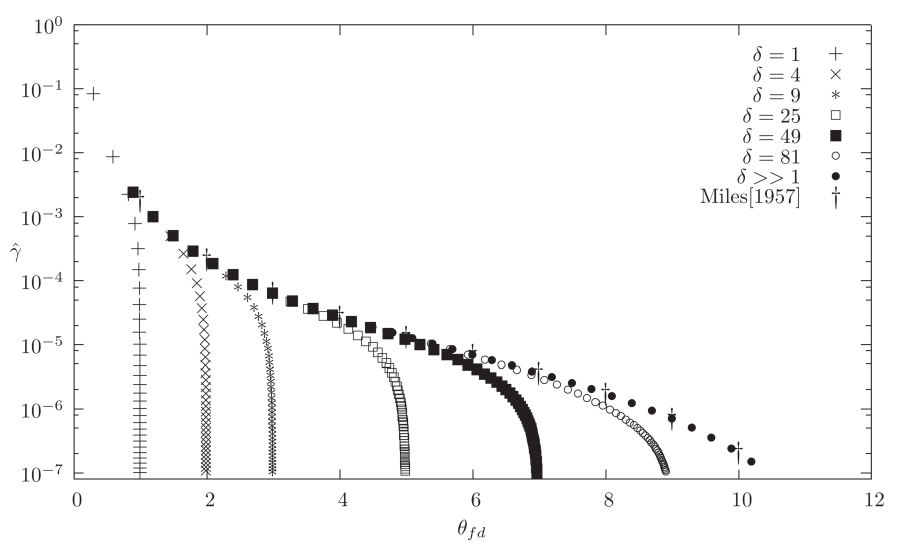}
	\caption{Evolution of the growth rate in semi-logarithmic scale. Every curve but the rightmost one correspond to finite depth. From left to right,  they match $\delta =1,4,9,25,49,81$. We can observe that for each depth, 
		there is a $\theta_{fd}$ -- limited wave growth. The deep water limit, also computed, is corresponds to small $\theta_{fd}$ and matches Miles' results.}
	\label{fig_growth_rate}
\end{figure}

The existence of a finite depth $h$
transforms the unique curve of wave growth rate in deep water in a \textit{family of
	curves} indexed by $\delta = gh/U^2_1$, i.e., a curve for each value of $\delta$.
In Figure \ref{fig_growth_rate} shows a family of six values of $\delta$ 
against the $\theta_{fd}$ parameter. The limit $ \delta \rightarrow \infty$ is included as well.
Small finite $\theta_{fd}$ corresponds to short surface waves. This stage represents the initial 
growth of the wave field near the shoreline of a calm lake. 
As time proceeds, the surface waves reach moderate $\theta_{fd}$, corresponding to mild or 
moderate wavelengths, while long waves are found for large $\theta_{fd}$. Of course, as the
wavelengths increase, the amplitudes keep on growing. Clearly, from a physical point of view, this means that 
Figure \ref{fig_growth_rate} is a snapshot of the theoretical dynamical development of the wave, which is growing in 
amplitude and wavelength in time.

Figure \ref{fig_growth_rate} shows that at small $\theta_{fd}$ the growth
rate $\gamma$ is equal for all values of $\delta$, the limit being the deep water case.
The finite-depth effects appear as $\theta_{fd}$ increases. The growth rate becomes lower than in the deep water 
limit for each value of $\delta$. The growth rates are scaled with $\delta$: for a given $\theta_{fd}$, the bigger the 
$\delta$ the larger the $\hat{\gamma}$. Each $\delta$-curve approaches its own (idealized) \emph{theoretical 
	$\theta_{fd}$-limited growth} as $\hat{\gamma}$
goes to zero. At this stage, the wave reaches a final linear progressive wave with zero 
growth. In other words, for a given $\delta$ the surface wave does not grow 
old beyond a determined $\theta_{fd}$.

\begin{figure}
	\centering
	\includegraphics[width=0.7\linewidth]{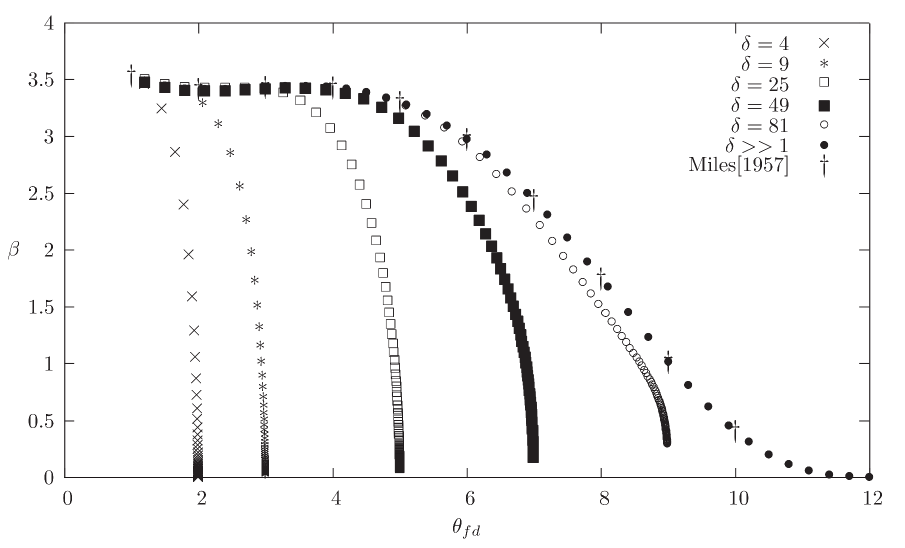}
	\caption{Evolution of Miles' coefficient $\beta$ for several values of the depth. Each curve is plotted with the same Charnock constant 
		$\alpha_c \approx 0.018$. The finite-depth effect is critical, and high value of $\delta$ correspond to deep water.}
	\label{fig_growth_rate_beta}
\end{figure}

In contrast to the usual analysis of wind-wave growth, our results concern the dimensionless growth rate
$\hat{\gamma}$ instead of the $\beta$-Miles parameter.
We have the following transformation rule between this parameter $\beta$ and
dimensionless $\hat{\gamma}$ 
\begin{equation}\label{betafinitedepth}
	\beta=\frac{2\hat{\gamma}}{s}\theta_{dw}^{3} T^{1/2},
\end{equation} 
where we took $\beta$ as it is usually defined, with the dimensions,
$\Im(c) = c_{0} \frac{s}{2} \beta (\frac{U_{1}}{c_{0}})^{2}.$
This definition of Miles' $\beta$ in finite depth is straightforward. Its evolution is shown clearly in Figure
\ref{fig_growth_rate_beta}, showing the correct deep water trends and the new finite depth limits. The effects of
depth are critical. As usual, $\beta$ is almost constant for small $\theta_{fd}$, but it goes dramatically to zero when the depth limit is close.
%%%%%%%%%%%%%%%%%%%%%%%%%%%%%%ine%%%%%%%%%%%%%%%%%%%%%%%%%%%%%%%%%%%%%%%%%%%%%%%%%%%%%%%%%%%%%%%%%%%%%%%%%%%%%%%%%%%%%%%%%%%%%%%%%%%%%%%%%%%%%%%%%%%%%%%

\subsection{Field experiments on growth of surface wind-waves}
\label{youngexperiments}
For finite depth wave growth
the pioneer experiments and numerical studies were conducted by 
\cite{IjimaTang, Thijsse, Bretschneider}
and particularly the experiments in Lake George, Australia, described by  
\cite{YoungVerhagen1}. They provided one of the first systematic attempts to 
understand the physics of wave-wind generation in finite-depth water. 

The results of the field experiments in fetch limited growth have been presented 
in references \cite{YoungVerhagen1} and \cite{YoungVerhagen2}. 
These papers described the basin geometry and 
bathymetry, experimental design, used instrumentation, and the adopted scaling parameters. The measurements have 
confirmed the water depth dependence of the asymptotic limits to wave growth. 

In reference \cite{Young1} (see also \cite{Young1999-zm})  derived an empirical relation in terms of 
appropriate dimensionless parameters able to reproduce the experimental data of \cite{YoungVerhagen1}. In particular, the empirical 
relationship between the fractional energy increase as a function of the inverse wave age, found by \cite{Donelan}
For deep water, it was extended to the finite depth domain.
Experimental results and plots of the empirical laws 
have shown that contrary to the deep water case, the wave age at which the growth rate becomes zero is wind-dependent and 
depth-dependent. So, the point of full development is warped from the deep water case, where established it \cite{PiersonMo}.
As a result, a growth law against the inverse wave age exists for each value of a parameter, which unites the dependences on wind intensity 
and water depth.

The evolution of the growth rates is such that at small wave ages, growth rates are comparable to the deep water limit at 
large wave ages, the growth rate is lower 
the growth rate vanishes beyond a limit wave age in shallow water than in deep water.
\subsection{Comparisons with field experiments}
\label{comparisons}
In this subsection, we are going to show that our analytical and numerical results can reproduce qualitatively 
these experimental facts.
At this point, it is essential to keep in mind that we are studying the linear growth of a normal 
Fourier mode $k$ and not the growth of a wave train as the infinite superposition of wave Fourier modes.

Moreover, results in field or laboratory experiments are commonly given using the parameter $C_{p}$,
the observed phase speed at the peak frequency $\omega_p$. Consequently, qualitative comparison with field 
observations can only be made using the phase velocity $c$ or frequency $\omega$ of one mode instead of $C_p$ or
$\omega_p$.

We are going to show that the theoretical curves for $\hat{\gamma}$ are, \textit{mutatis mutandi}, in good qualitative agreement with the empirical curves of the dimensionless fractional wave energy increase per radian $\hat{\Gamma}$  as a function of the inverse wave age $U_{10}/Cp$ in \cite{Young1}. In this reference, experimental field data for $\hat{\Gamma}$ in the finite depth of the empirical relationship adequately represents Lake George
\begin{eqnarray}
	\hat{\Gamma}&=&\frac{C_g}{\omega_p} \frac{1}{E}\frac{\partial E}{\partial x}\\&=&
	A\left(\frac{U_{10}}{C_p}-0,83\right)\tanh^{0,45}{\left(\frac{U_{10}}{C_p}-\frac{1,25}
		{\delta^{0,45}_Y}\right)}\nonumber,
\end{eqnarray}
with A constant, $\delta_Y=gh/U_{10}^2$ the non-dimensional water depth, $U_{10}$ the wind velocity at 
$10~m$, and $C_g$ and $C_p$ the group and phase speeds of the components at the spectral peak frequency $\omega_p$.

\begin{figure}
	\begin{tabular}{c}
		\includegraphics[width=0.48\linewidth]{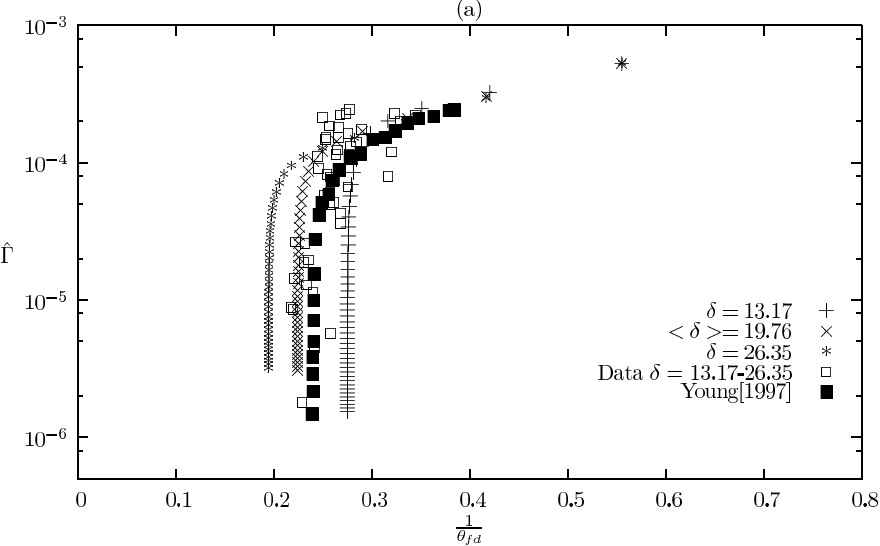} \hskip0.01cm
		\includegraphics[width=0.48\linewidth]{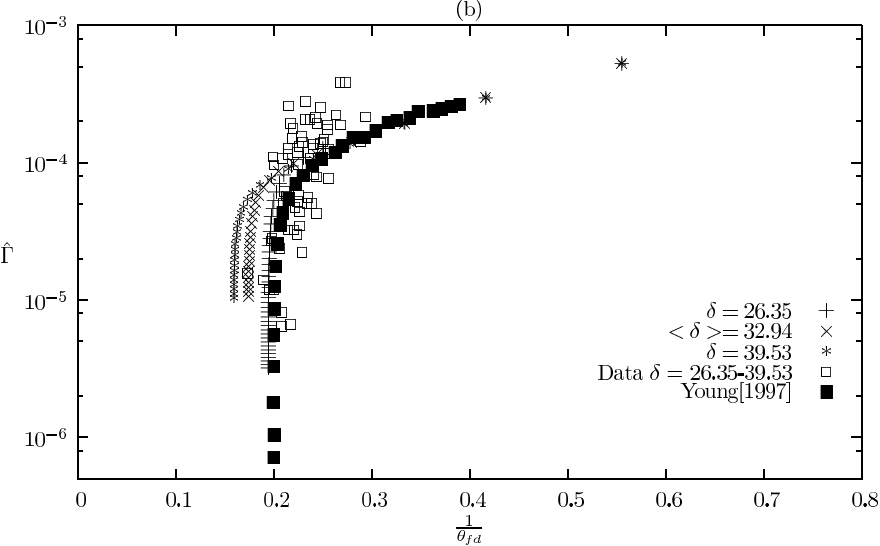}\\
		\includegraphics[width=0.48\linewidth]{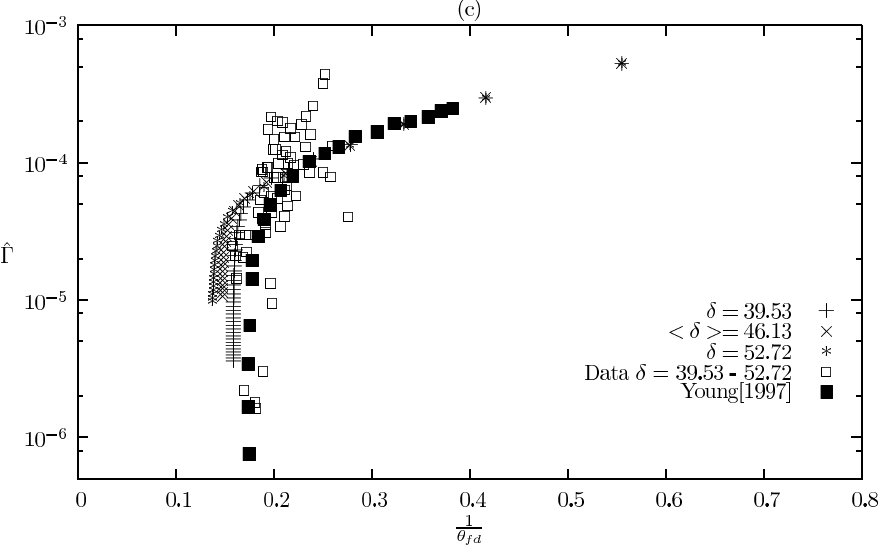}\hskip0.01cm
		\includegraphics[width=0.48\linewidth]{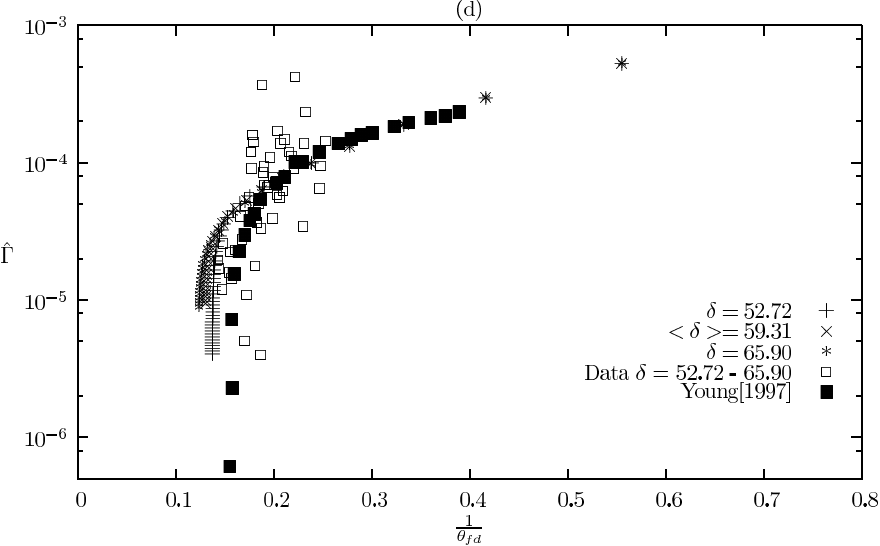}\\
	\end{tabular}
	\caption{Growth rate $\hat{\Gamma}$ as a function of inverse wave age $1/\theta_{fd}$
		for several values of the parameter $\delta$. White squares correspond to Lake George
		experiment data, Black squares correspond to the empirical relationship (eq. (6)) in
		\protect\cite{Young1}. Present results correspond to symbols $+$, $\times$ and $\ast$. a: the
		dataset covers a range of wind speed corresponding to $\delta_Y = 0.1-0.2$, or using
		(\ref{convertdelta}) $\delta = 13.17-26.35$, and an average value $
		<\delta>=(13.17+26.35)/2$ is used. b: same as \lq\lq{}a\rq\rq{} with $\delta_Y = 0.2-0.3$. c: same
		as \lq\lq{}a\rq\rq{} with $\delta_Y = 0.3-0.4$. d : same as \lq\lq{}a\rq\rq{} with $\delta_Y = 0.4-0.5$.}
	\label{fig_plotyoung}
\end{figure}

In order to make a qualitative comparison between $\hat{\Gamma}$ curves in function of the inverse wave-age $U_{10}/C_p$
and theoretical $\hat{\gamma}$ curves in function of $1/\theta_{fd}$ we need  
to write the empirical $\hat{\Gamma}$ in terms of theoretical quantities.
So, the following changes are necessary:
\begin{eqnarray}
	\mbox{measured}& C_g, C_p, \omega_p \rightarrow \mbox{theoretical} \hskip2mmc_g, c, \omega,\label{C2c}\\
	\mbox{and }&\quad \frac{U_{10}C_{10}^{1/2}}{\kappa} = u_{*}/\kappa =U_{1} \label{U1O}
\end{eqnarray}
with $C_{10}$ the $10~m$ drag coefficient \cite{JinWu}. Thus, from the fact that the energy growth rate is two times the amplitude growth rate, that is
$$
\Gamma=2\gamma,
$$
and using $2c_g=c(1+2kh/\sinh(2kh))$, (\ref{C2c}), (\ref{U1O}),(\ref{waveagefinite}) and (\ref{parameters}) we obtain
\begin{equation}
	\hat{\Gamma}=\frac{\theta_{dw}}{T^{1/2}}\hat{\gamma}[1+\frac{2\delta}{\theta_{dw}^2\sinh(\frac{2\delta}{\theta_{dw}^2})}].
\end{equation}
This expression gives the theoretical equivalent of the empirical $\hat{\Gamma}$ in function of $\theta_{dw}$, $\delta$
and $\hat{\gamma}$.
The values of $\hat{\gamma}$ for fixed $\delta$'s as a function of $1/\theta_{fd}$ are numerically obtained from
Eqs. (\ref{waveagefinite}) (\ref{tanh}) and (\ref{growthrate}).
Steps (\ref{C2c}) and (\ref{U1O}) transform $\delta_Y$ and $C_p/U_{10}$ into $\delta$ and $\theta_{fd}$ according to
{\begin{eqnarray}
		\delta_Y &=&\delta\frac{C_{10}}{\kappa^2}\label{convertdelta},\\
		\frac{C_p}{U_{10}} &=& \theta_{fd}\frac{C_{10}^2}{\kappa}.
\end{eqnarray}}
In reference \cite{Young1} the curves of $\Gamma$ versus $U_{10}/C_p$ have been presented for the $\delta_Y$-intervals $\delta_Y \in [0.1-0.2],[0.2-0.3],
[0.3-0.4],[0.4-0.5]$, rather than for a single value of $\delta_Y$. The intervals were 
determined from the variations in $U_{10}$, the depth $h$ being nearly constant around $2~m$. 
Consequently we substitute the $\delta_Y$-intervals with $\delta$-intervals using (\ref{convertdelta}) and we evaluate the mean value $\overline{\delta}$.
For example $\delta_Y \in [0,1-0,2]$ is transformed into $\delta=[13,17-26,35]$ with
$\overline{\delta}=19,76$ in Figure \ref{fig_plotyoung}(a).
Figures \ref{fig_plotyoung}(a),  \ref{fig_plotyoung}(b),  \ref{fig_plotyoung}(c) and  \ref{fig_plotyoung}(d) are displaying a 
fair concordance of the model with the experimental data and plots of empirical laws
for Lake George. The agreement improves as $\frac{1}{\theta_{fd}}$ increases.

\begin{figure}
	\centering
	\includegraphics [width=0.7\linewidth]{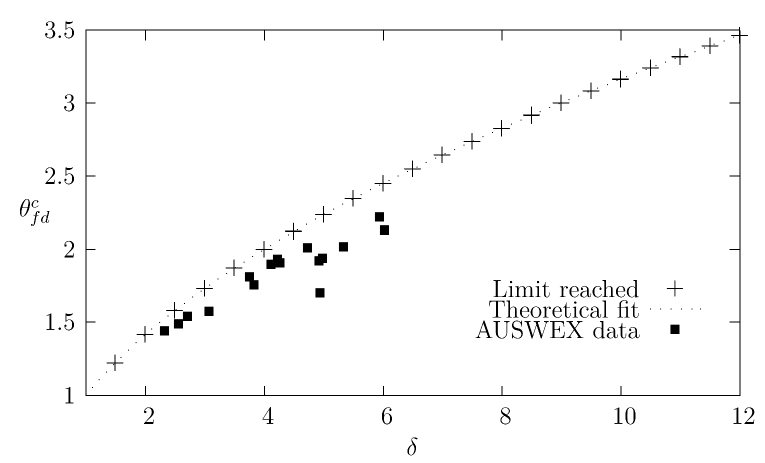}
	\caption{Parameter curves corresponding to zero growth rate. The theoretical limit is given
		by (\ref{thetacritique}). The AUSWEX data are experimental results from \protect\cite{Donelan}
		(the sea state is fairly close to the finite depth full development).}
	\label{gamma_zero}
\end{figure}

In Figure \ref{gamma_zero} are plotted, against $\delta$, the critical values of the parameter $\theta_{fd}^{c}$ for which 
the growth rate $\gamma$ goes to zero. They obey the relation
\begin{equation}
	\label{thetacritique}
	\theta_{fd}^{c} =\delta^{0,5}.
\end{equation}
The above relation, found numerically, is coherent with the parameter formulation (\ref{waveagefinite}). 
It is indeed a limiting value for $\theta_{fd}$ uniquely determined by the water depth. 
In \cite{Young1} the author has shown 
from an empirical relationship (formula (6) in reference above) that $\hat{\Gamma}$ the growth rate goes to zero 
as a function of the inverse wave age $ U_{10}/C_p $ for
\begin{equation}\label{Gammazero}
	\frac{C_p}{U_{10}}=0.8(\frac{gh}{U_{10}^{2}})^{0.45}.
\end{equation}
Using a $C_{10}$ drag coefficient parametrization such as \cite{JinWu}
\begin{equation}
	C_{10} = (0.065 U_{10} +0.8) 10^{-3},
\end{equation}
and taking an average $U_{10} = 7~m/s$ in \cite{Young1}, one finds the $U_1$ to $U_{10}$ relationship
\begin{equation}
	U_{10}\approx 28,3~u_{*}\approx11,6~U_1,
\end{equation}
So, this limiting law reads
\begin{equation}
	\frac{C_p}{U_{1}} = 1,01~\delta^{0,45}.
	\label{C10}
\end{equation}
a result in excellent agreement with the theoretical value (\ref{thetacritique}).
With $\theta_{fd}^{c}$ we can calculate the corresponding critical wave length $\lambda^{c}$. Using 
(\ref{thetacritique}) in (\ref{waveagefinite}) we obtain
\begin{equation}\label{shallow}
	\frac{\delta}{\theta^2_{dw}}=\tanh{(\frac{\delta}{\theta^2_{dw}})}.
\end{equation}
Relation (\ref{shallow}) means the wave has entered the shallow water region.
In such a limit the range of $\delta/\theta^2_{dw}$ is: $0<\delta/\theta^2_{dw}<\frac{\pi}{4}$ (\cite{Fenton, Marcus}). As a result we obtain $\lambda^{c} = 8 h$. For values of $\lambda$ such that $\lambda > \lambda^{c}$ 
the phase velocity is in the long wave limit i.e., $c=\sqrt{g h}$. Consequently, if $\lambda > \lambda_c$ the wave feels 
the bottom, the amplitude does not grow anymore, the resonance wind/phase speed ceases, and the wave reaches its utmost 
state as a progressive plane wave.

Finally in Figure \ref{gamma_zero} are also represented data from \cite{Donelan}, from the Australian Shallow Water 
Experiment. A fit is also plotted to show the trend.
The raw data consists in the water depth $h$ in meters, the friction velocity $u_{*}$, the $10$ meters wind 
velocity $U_{10}$ and the ratio of the former with the measured phase speed $c_p$, $U_{10}/c_p$. For example, 
$u_{*} = 0.44~m.s^{-1}$ and $h = 0.32~m$ gives $\delta = 2.7$ and 
$\theta_{fd} = 1.55$, which gives a small relative error regarding (\ref{thetacritique}). All the points 
give $(\delta,\theta_{fd})$ coordinates close to the theoretical limit. 
%%%%%%%%%%%%%%%%%%%%%%%%%%%%%%%%%%%%%%%%%%%%%%%%%%%%%%%%%%%%%%%%%%%%%%%%%%%%%%%%%%%%%%%%%%%%%%%%%%%

%%%%%%%%%%%%%%%%%%%%%%%

\section{A nonlinear Miles' approach} 

\label{Miles}
In this section we consider the air/water system from quasi-linear point of view in the sense that the water dynamics is considered nonlinear and irrotational but the airflow is kept linear . We use the $\beta$-Miles parameter as given by Eq. \ref{betafinitedepth}.
In Miles's theory of wave generation \cite{Miles1, Miles2}, the complex air pressure $P_a$ can be separated into two components, one 
in phase and one in quadrature with  the free surface $\eta$. A phase shift between those two quantities is necessary to transfer energy from the air flow to 
the wave field. The transfer is only due to the part of $P_a$ in quadrature with $\eta$. Hence, we will deal only with the acting pressure component, that is
\begin{equation}\label{Pa}
	P_a(x,t)= \rho_a \beta U_1^2 \eta_x(x,t),
\end{equation}
\subsection{The anti-diffusive nonlinear Schr\"odinger equation in finite depth}
\label{nonlinearNLS}
Let us consider the air/water system from a {\it quasi-linear} point of view i.e., the water dynamics is considered nonlinear and irrotational and, as in Miles' theory,
the air flow is kept linear. So with this assumption
the complete irrotational Euler equations and boundary conditions in terms of 
the velocity potential $\phi(x,z,t)$ are
\begin{eqnarray}
	\phi_{xx} + \phi_{zz} = 0 \quad \mbox{for} \quad -h \le z \le
	\eta,\label{laplace}\\
	\phi_z =0 \quad \mbox{for}\quad z=-h,\label{BClaplace}\\
	\eta_t + \phi_x\eta_x - \phi_z = 0
	\quad \mbox{for}\quad  z= \eta,\label{BCkine}\\
	\phi_t + \frac12\phi_x^2 + \frac12\phi_z^2 +g\eta=-\frac{1}{\rho_w}P_a
	\quad \mbox{for} \quad z=\eta.
\end{eqnarray} 
Using (\ref{Pa}) the  modified Bernoulli equation reads
\begin{equation}
	\phi_t + \frac12\phi_x^2 + \frac12\phi_z^2 +g\eta = -s
	\beta U_1^2\eta_x \quad \mbox{for}\quad z = \eta.\label{BCdyn}
\end{equation}
From equations (\ref{laplace}), (\ref{BClaplace}),
(\ref{BCkine}) and (\ref{BCdyn}) we find a wind-forced finite depth NLS equation
for $\eta$ as a function of the standard slow space and time variables $\xi=\varepsilon(x-c_g t)$ 
and $\nu=\varepsilon^2 t$, ($\varepsilon << 1$) and $c_g$ the group velocity. The perturbed NLS equation reads
\begin{equation}
	i\eta_{\nu} + a\eta_{\xi\xi}+b|\eta|^2\eta=id\eta
\end{equation}
with $c_g, a, b$ and $d$ given by
\begin{eqnarray}
	c_g&=&\frac{c}{2}[1+2kh/\sinh(2kh)],\nonumber\\
	a&=&-\frac{c_g^2 -gh[1-khT(1-T^2)]}{2\omega},\nonumber\\
	b&=&\frac{k^4c^2}{4\omega T^2}[\frac{9}{T^2}-12+13T^2 
	-2T^4-\frac{2[2c+c_g(1-T^2)]^2}{gh-c_g^2}],\nonumber\\
	d&=&s\frac{\beta}{2}\frac{U_1^2}{c^2}T\omega. \nonumber
\end{eqnarray}
For more information about the derivation of the coefficients $a$ and $b$ see \cite{Thomas}. To derive a dimensionless 
wind-forced NLS equation we use (\ref{parameters}) and we obtain
in the original laboratory variables $x$ and $t$ (after a Galilean transformation in order to eliminate the linear term
$c_g \eta_{x}$ and dropping the hats)
\begin{equation}\label{NLSdimensionless}
	i\eta_t + A\eta_{xx}+B|\eta|^2\eta=iD\eta
\end{equation}
with $c_g, A, B$, and $D$ now given by
\begin{eqnarray}\label{}
	c_g&=&\frac{1}{2\theta_{fd}}[1+\frac{\delta}{\theta_{dw}^2}\frac{1-T^2}{T}],\nonumber\\
	A&=&-\frac{c_g^2 -\delta[1-\delta\theta_{dw}^{-2}(1-T^2)]}{2\theta_{fd}\theta_{dw}^{2}},\nonumber\\
	B&=&\frac{1}{4T^2\theta_{fd}^3\theta_{dw}^{2}}[\frac{9}{T^2}-12+13T^2
	-2T^4-\frac{2[2\theta_{fd}^{-1}+c_g(1-T^2)]^2}{\delta-c_g^2}],\nonumber\\
	D&=&s\frac{\beta}{2}
	\frac{T^{1/2}}{\theta_{dw}^{3}}.\nonumber
\end{eqnarray}

Eq. (\ref{NLSdimensionless}) is a wind-forced finite depth NLS equation in dimensionless variables.
\subsection{The Akhmediev, Peregrine and Ma solutions for weak wind inputs
	in finite depth}
\label{Freakwavessolution}

The wind-forced nonlinear Schr\"odinger equation allows the study of the wind influence 
on the freak waves dynamics \cite{Touboul2,Touboul1,Kharif,Onorato}.
Previous authors have carried out such studies in deep water. 
The present work allows,  similar studies in finite depth
with the right Miles' growth rates. In the following we are going only to consider the so called focusing NLS equation i.e., positive $A$ and $B$.
Introducing $\eta'$ and $x'$ as 
$$
\eta'=\sqrt{B}\eta, \quad  x'=\frac{x}{\sqrt{A}},
$$
Dropping the primes, Eq. (\ref{NLSdimensionless}) becomes
\begin{equation}\label{NLSdimensionless2}
	i\eta_t + \eta_{xx}+|\eta|^2\eta=iD\eta.
\end{equation}
Introducing a function $M(x,t)$ as
\begin{equation}
	M(x,t)=\eta(x,t)\exp{(-Dt)},
\end{equation}
we obtain from (\ref{NLSdimensionless2})
\begin{equation}\label{NLSdimensionless3}
	iM_t+M_{xx}+\exp{(2Dt)}|M|^2M=0.
\end{equation}

In order to reduce Eq. (\ref{NLSdimensionless2}) into the standard form of the NLS with constant coefficients 
we proceed in the following way. First of all we consider the wind forcing $(2Dt)$ to be weak, such that the exponential 
can be approximated by $\exp{(2Dt)} \sim 1+2Dt$ so 
\begin{equation}\label{NLSdimensionless4}
	iM_t+M_{xx}+n|M|^2M=0,\quad n=n(t)= 1-2Dt.
\end{equation}

Now with a change of coordinates from $(x,t)$ to $(z,\tau)$ defined by
\begin{equation}
	z(x,t)=xn(t),\quad \tau(x,t)=xn(t),
\end{equation}
and scaling the wave envelope as \cite{Onorato}
\begin{equation}
	M(z,\tau)=\Psi(z,\tau)\sqrt{n(\tau)}\exp{(\frac{-iDz^2}{n(\tau)})},
\end{equation}
we reduce (\ref{NLSdimensionless4}) to the standard focusing equation
\begin{equation}\label{NLSforPsi}
	i\Psi_{\tau} + \Psi_{xx} + |\Psi|^2\Psi=0.
\end{equation}
Equation (\ref{NLSforPsi}) admits well known breather solutions that are simple analytical
prototypes for rogue wave events. They are the Akhmediev ($\Psi_A$) \cite{Akhmediev}, the Peregrine ($\Psi_P $) \cite{Peregrine} and the Kuznetsov-Ma ($\Psi_M $) \cite{Ma} breather solutions. 
\cite{Dysthe} investigated whether freak waves in deep water could be modeled by $\Psi_A$, $\Psi_P $ or by $\Psi_M $. \cite{Onorato} considered the influence of weak wind forcing and dissipation on these
$\Psi_A$, $\Psi_P $ or $\Psi_M $ solutions in deep water.
The present work allows us to exhibit expressions
for $\Psi_A$, $\Psi_P$ and $\Psi_M $ {\it under the influence of weak wind forcing in finite depth $h$ given by the extended Miles mechanism.}
These solutions read \cite{Dysthe}:
\begin{equation}
	\eta_A=P(\tau)\{\frac{\cosh(\Omega\tau-2i\omega)-\cos(\omega)\cos(pz)}{ \cosh(\Omega\tau)-\cos(\omega)\cos(pz)}\},
\end{equation}
with $p=2\sin(\omega)$, $\Omega=2\sin(2\omega)$  $\omega$ real and $p$ related to the spatial period $2\pi/p$
\begin{equation}
	\eta_P=P(\tau)\{1-\frac{4(1+4i\tau)}{1+4z^2+16\tau^2}\},
\end{equation}
\begin{equation}
	\eta_M=P(\tau)\{\frac{\cos(\Omega\tau-2i\omega)-\cosh(\omega)\cosh(pz)}{ \cosh(\Omega\tau)-\cos(\omega)\cos(pz)}\},
\end{equation}

where $p=2\sinh(\omega)$, $\Omega=2\sinh(2\omega)$ and $\Omega$ real and related to the time period $2\pi/\Omega$
and
$$
P(\tau)=n(\tau)\exp{[\frac{-iDz^2}{n(\tau)}]}\exp{[2i\tau]}.
$$

 It is worth noticing that many works are done on Akhmediev solutions and rogue
	waves. We can mention here one of the recent works of Grinevich and Santini \cite{Grinevich_2021} and references therein.

 \section{The Nonlinear Jeffrey's approach} 
 
 \label{Jeffreys}
 
\begin{figure}
 	\centering
 	\includegraphics[width=0.6\linewidth]{"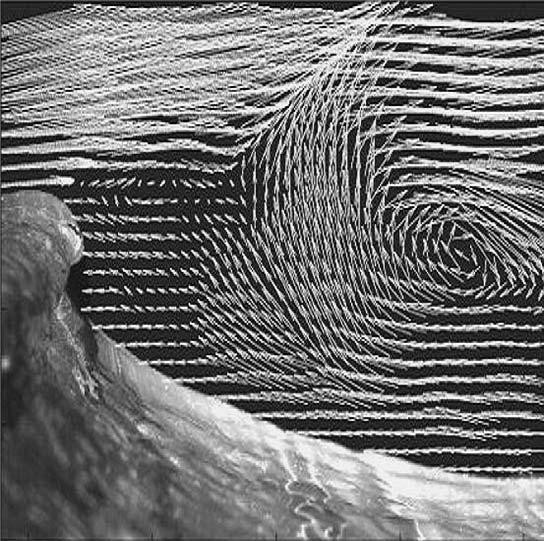"}
 	\caption{Jeffreys' mechanism  - Airflow separation observed by "Particle Image Velocimetry" over a breaking wave \cite{10.1063/1.870058}.	The air pressure on the windward face of the wave is larger than the leeward face, which is the origin of continuous energy transfer from wind to wave.  }
 	\label{fig:jeffrys-mechanism}
 \end{figure}

 Jeffreys introduced a second significant mechanism, extensively researched today, in 1925. The principle was that waves if sufficiently disrupted, could break the continuity of the airflow and induce a gradient of pressure. This understanding could have practical implications, allowing the wave to grow quickly according to the relative distance between the phase speed of the wave and the wind speed.
 Jeffreys found the main idea behind this mechanism by analogizing it with a sphere immersed in a laminar fluid flow. Jeffreys noticed that, in the viscous case, the fluid particles do not do everything around the sphere. Those hitting the forehead slide on the sphere and take off shortly after the halfway point, and on the other side, an area is almost stagnant and has a relative speed relative to the sphere close to zero. (1925)
 When a wave is large enough, it breaks the continuity of the airflow, and a horizontal axis vortex forms on the side sheltered from the wind by the wave. Other smaller whirlpools form between the large vortex and the laminar flow. This phenomenon, called airflow separation, has been observed in waves, see Figure \eqref{fig:jeffrys-mechanism}. In this case, the face of the wave opposing the wind will experience greater pressure, and the sheltered side will experience less pressure. This gradient induces a reaction at the surface level.

 %%%%%%%%%%%%%%%%%%%%%%%%%%%%%%%%%%%%%%%%%%%%%%%%%%%%%%%%%%%%%%%%%

 %%%%%%%%%%%%%%%%%%%%%%%%%%%%%%%%%%%%%%%%%%%%%%%%%%%%%%%%%%%%%%%%%%%%%%%%%
 
 \subsection{The Nonlinear Serre-Green-Naghdi Approximation}\label{2-1}
 We associated water particles, in a system of two-dimensional Cartesian coordinates $(x,z)$ with origin $0$, where $z$ is the upward vertical direction. We let 
 $z = 0$ at the water--air interface. Hence, the positive values of $ z $, {$z \in~]0, \infty[$}, % It is correct.
 correspond to the (unperturbed) air domain, while negative values of $ z $, $z\in [-h, 0]$ correspond to the (unperturbed) water domain. Consequently, for the bottom of the water domain of depth $ h $, we obtain $z = -h$. The bottom is considered to be impermeable, and both water and air, are taken to be inviscid and incompressible. Moreover, the surface tension effects, at the interface, are not taken into account. The governing equations are the well known Euler equations, with the mass conservation equations used in $(x,z,t)$ frame, where $ t $ accounts for the {time. Namely,} %MDPI: Sub equation number like 1a, 1b, 1c is not suggested. Please number the equations in numerical order (1, 2, 3) instead of using a,b,c.
 \begin{subequations}\label{EulerWater}
 	\begin{align}
 		&u_x + w_z = 0, \label{cont}\\
 		& \rho_w (u_t+uu_x+wu_z) = -P_{x}, \label{Eulerx}\\
 		& \rho_w( w_t + uw_x + ww_z)=-P_{z}-g\rho_w, \label{Eulerz}
 	\end{align}
 \end{subequations}
 where $u(x,z,t)$ and $w(x,z,t)$ are the fluid's horizontal and vertical velocities, respectively. $P(x,z,t)$ is the Archimedean pressure, $g$ the gravitational acceleration, $\rho_w$ is the water density and
 subscripts $x$, $z$ and $t$ denote partial derivatives with respect to $x$, $z$ and $t$, respectively. The set of Equations \eqref{EulerWater} are completed by the following boundary conditions:
 %The boundary conditions at $z = -h $ and at the perturbed water surface $z = \eta (x,t)$ are
 \begin{subequations}
 	\begin{align}
 		&w= 0,\quad\text{at}\quad z=-h, \label{bottonBC0}\\
 		&\eta_{t}+u\eta_x-w=0,\quad\text{at}\quad z=\eta, \label{EulerWaterBC1}\\
 		&P= P_a,\quad\text{at}\quad z=\eta, \label{EulerWaterBC2}
 	\end{align}
 \end{subequations}
 where $P_a(x,z,t)$ is the air pressure, and Equation (\ref{EulerWaterBC2}) expresses the pressure's continuity across the air/water interface. Notice that $z = \eta (x,t)$ is the perturbed water surface. For convenience, we introduced a reduced pressure $ \mathbf{P} $, such that 
 $ \mathbf{P}(x,z,t)=P(x,z,t) + \rho_w gz -P_{0 }$,
 where $P_0$ denotes the atmospheric pressure. Using the reduced pressure $\mathbf{P}(x,z,t)$, the set of Equations \eqref{EulerWater}
 can be written as follows:
 \begin{subequations}
 	\begin{align}
 		%u_x + w_z &=& 0,\label{cont2} \\
 		& \rho_w (u_t+uu_x+wu_z) = -\mathbf{P}_x,\label{Eulerx2} \\
 		& \rho_w( w_t + uw_x + ww_z)=-\mathbf{P}_z,\label{Eulerz2}\\
 		&\mathbf{P}(x,\eta,t)-\rho_w g \eta+ P_{0} = P_a(x,\eta,t). \label{EulerWaterBC22}
 	\end{align}
 \end{subequations}
 
 {Shallow water model equations,} %MDPI: We added indentation here. Please confirm. Please confirm all the indentation of the paragraph after equations in this paper.
 such as the Korteweg-Vries, modified Korteweg-Vries, and Boussinesq equations and many others, are usually derived by performing an asymptotic analysis directly from the equations of continuity (\ref{cont}), the motion Equations (3a,b) and the boundary conditions (2a,b) and (\ref{EulerWaterBC22}) (see \cite{JAFD_2017} and references therein).
 In this work, our approach was somewhat different, in the sense that instead of applying a perturbation theory to the entire problem, we first considered the nonlinear evolution of a given velocity
 field profile. Indeed, we assumed the horizontal velocity $u(x,z,t)$ to be independent of $z$, i.e., 
 \begin{equation}\label{Anzats}
 	u=u(x,t).
 \end{equation}
 
 The choice of Equation \eqref{Anzats}, is known as the \textit{columnar flow hypothesis}, which was introduced by \cite{Su}, and \cite{Serre}.~Using Equations (\ref{cont}), (\ref{bottonBC0}) and (\ref{Anzats}) we~obtain 
 \begin{equation}
 	w(x,z,t)=-(z+h)u_x(x,t).
 \end{equation}
 
 Hence, Equations (\ref{EulerWaterBC1}) and (3a,b) read
 \begin{subequations}
 	\begin{align}
 		&\eta_t+ [u(\eta + h) ]_x=0,\label{GN2}\\
 		& \rho_w(u_t+uu_x)=-\mathbf{P}_x, \label{Eulerx3}\\
 		&\rho_w(z+h)(u_{xt}+uu_{xx}-u^2_x)=\mathbf{P}_z.\label{Eulerz3}
 		% \eta_t+ [u(\eta + h) ]_x&=&0.\label{GN2}
 	\end{align}
 \end{subequations}
 
 The integration of Equation (\ref{Eulerz3}), using Equation (\ref{EulerWaterBC22}), yields the pressure $\mathbf{P}(x,z,t)$ :
 \begin{eqnarray}
 	\mathbf{P}(x,z,t)=\frac{1}{2}\rho_w[(z+h)^2-(\eta+h)^2] (u_{xt}+uu_{xx}-u^2_x)
 	+ P_a(x,\eta,t) + \rho_w g\eta-P_0 \label{Eulerz3int}.
 \end{eqnarray}
 
 The next step consists of substituting Equation (\ref{Eulerz3int}) in Equation (\ref{Eulerx3}), and taking the $z$-average of Equation (\ref{Eulerx3}) for $-h\leq z\leq \eta$. 
 Finally, using Equation (\ref{EulerWaterBC22}) we obtained the following~system:
 \begin{subequations}\label{GN}
 	\begin{align}
 		& \eta_t+ [u(\eta + h) ]_x =0,\label{GN1}\\
 		& u_{t}+uu_x+g\eta_x -\frac{1}{3(\eta+h)}\{(\eta+h)^3(u_{xt}+uu_{xx}-u^2_x)\}_x=-\frac{1}{\rho_w}[P_a(x,\eta,t]_{x}. \label{GN2}  
 	\end{align}
 \end{subequations}
 
 If $P_a=P_0$, Equations (8a,b) are reduced to the usual Serre--Green--Naghdi equations. However, in our approach, $P_a$ was not taken as equal to $P_0$, and the expression of $P_a(x,z,t)$ was found using the sheltering mechanism.{\color{red}{}}
 
 \subsection{Jeffreys' Sheltering Mechanism of Wind Waves Generation Applied to Serre-Green-Nagdhi~Equations} \label{water domain}

 The physical sheltering mechanism assumes that the energy transfer is caused by pressure drag (also known as ``form drag'').
 The air pressure on the windward face of the wave is larger than the leeward face, which is the origin of continuous energy transfer from wind to wave. 
 Using dimensional arguments, Jeffreys \cite{Jeffreys1,Jeffreys2} showed that the air pressure perturbation 
 $P_{a}(x,z,t)$
 evaluated on the surface can be represented by
 \begin{equation}\label{Jeffreys}
 	P_{a}(x,z=\eta,t)=\rho_{a}\epsilon (U_{10}-c)^{2}\eta_{x}(x,t),
 \end{equation}
 where $\epsilon$ is the {\it sheltering coefficient}, $\rho_{a}$ is the air density and $U_{10}$ is the wind velocity at a 10 m height. 
 The sheltering coefficient is less than $1$ $(\epsilon<1)$. By substituting Equation (\ref{Jeffreys}) in Equation (\ref{GN2}), we obtained 
 \begin{subequations}\label{GNWind}
 	\begin{align}
 		&\eta_t+ [u(\eta + h) ]_x=0, \label{GNW1}\\
 		& u_{t}+uu_x+g\eta_x -\frac{1}{3(\eta+h)}\{(\eta+h)^3(u_{xt}+uu_{xx}-u^2_x)\}_x  =-\epsilon s \Delta^{2}\eta_{xx},\label{GNW2} 
 	\end{align}
 \end{subequations}
 where $s=\frac{\rho_{a}}{\rho_{w}}\sim 10^{-3}$ and $\Delta=(U_{10}-c).$ Thus, Equation \eqref{GNWind} constitutes the fully nonlinear Serre--Green--Naghdi system describing surface 
 wave propagation in shallow water under the action of the wind sheltering mechanism.
 
For convenience, we introduced new variables $S(x,t)$, $U(x,t)$ and $\alpha$, defined as follows:
 \begin{subequations}\label{newvariables}
 	\begin{align}
 		& S(x,t)=\eta(x,t)+h \label{newvariables1},\\
 		& U(x,t)=u(x,t)\label{newvariables2},\\
 		&\alpha=\epsilon \Delta^{2}\label{newvariables3}.
 	\end{align}
 \end{subequations}
 
 Using the variables \eqref{newvariables}, the system of Equation (10a,b) can be written as follows:
 \begin{subequations}\label{US}
 	\begin{align}
 		&	S_{t}+U_{x}S+US_{x}=0,\label{US1}\\
 		& U_{t}+UU_{x}+gS_{x}-\dfrac{1}{3S}\bigg\{S^{3}\big[U_{xt}+UU_{xx}-(U_{x})^{2}\big]\bigg\}_{x}=-\alpha sS_{xx}.\label{US2}
 	\end{align}
 \end{subequations}
 
 Considering the following frame $ \sigma $, and the slow time $ \tau $:
 \begin{subequations}\label{change of variable}
 	\begin{align}
 		&	\sigma = x-vt,\\
 		& \tau =st,
 	\end{align}
 \end{subequations}
 and applying Leibniz's chain rules by considering the change of variables \eqref{change of variable}, the $ x $ and $ t $ derivatives can be expressed as follows:
 \begin{equation}\label{chain rule}
 	\partial_{x}=\partial_{\sigma}, \quad \partial_{t}=-v\partial_{\sigma}+s\partial_{\tau}, \quad \partial^{2}_{xt}=-v\partial^{2}_{\sigma\sigma}+s\partial^{2}_{\tau\sigma}, \quad \partial_{xx}=\partial_{\sigma\sigma},\quad  \partial_{xxx}=\partial_{\sigma\sigma\sigma}.
 \end{equation}
 
 With the parameter $s$ being small ($s\sim 10^{-3}$), Equation (12a,b) can be expanded in terms of $ s $, as follows:
 \begin{subequations}\label{expansion}
 	\begin{align}
 		&	U=U_{0}+sU_{1}+O(s^{2}),\\
 		& 	S=S_{0}+sS_{1}+O(s^{2}).
 	\end{align}
 \end{subequations}
 
 Notice that the expansion should not be continued beyond the first order, because Equation (\ref{US2}) is an $ s^{1} $-order equation. Using the partial derivate chain rules \eqref{chain rule} in the SGN Equation \eqref{US} yields the following: \\
 
 \noindent {{At order 0}} %MDPI: Please check if the bold is necessary here and if the noindent format is correct here. Please check all the contents in this format.
 \begin{subequations}\label{GN order 0}
 	\begin{align}
 		&	-vS_{0,\sigma}+U_{0,\sigma}S_{0}+U_{0}S_{0,\sigma}=0,\label{eq:GN order 0-1}\\
 		&
 		-vU_{0,\sigma}+U_{0}U_{0,\sigma}+gS_{0,\sigma}+S_{0}S_{0,\sigma}\big[vU_{0,\sigma\sigma}-U_{0}U_{0,\sigma\sigma}+(U_{0,\sigma})^{2}\big] \nonumber\\
 		&
 		\hskip4cm+\dfrac{1}{3}(S_{0})^{2}\big[vU_{0,\sigma\sigma\sigma}-U_{0}U_{0,\sigma\sigma\sigma}+U_{0,\sigma}U_{0,\sigma\sigma}\big]=0.\label{eq:GN order 0-2}
 	\end{align}
 \end{subequations}
 {{At order 1}} 
 \begin{subequations}\label{GN order 1}
 	\small{\begin{align}
 			&-vS_{1,\sigma}+U_{0,\sigma}S_{1}+U_{1,\sigma}S_{0}+U_{0}S_{1,\sigma}+U_{1}S_{0,\sigma}=-S_{0,\tau},\\
 			&-vU_{1,\sigma}+U_{0}U_{1,\sigma}+U_{1}U_{0,\sigma}+gS_{1,\sigma}+(S_{0}S_{1,\sigma}+S_{1}S_{0,\sigma})\left[vU_{0,\sigma\sigma}+(U_{0,\sigma})^{2}-U_{0}U_{0,\sigma\sigma}\right]
 			\nonumber \\	&+S_{0}S_{0,\sigma}\left[vU_{1,\sigma\sigma}-U_{0}U_{1,\sigma\sigma}-U_{1}U_{0,\sigma\sigma}+2U_{0,\sigma}U_{1,\sigma}\right]
 			\nonumber \\	
 			&+\dfrac{1}{3}(S_{0})^{2}\big[vU_{1,\sigma\sigma\sigma}-U_{0}U_{1,\sigma\sigma\sigma}-U_{1}U_{0,\sigma\sigma\sigma}+U_{0,\sigma}U_{1,\sigma\sigma}+U_{1,\sigma}U_{0,\sigma\sigma}\big]
 			\nonumber\\
 			&+\dfrac{2}{3}S_{1}S_{0}\left[vU_{0,\sigma\sigma\sigma}+U_{0,\sigma}U_{0,\sigma\sigma}-U_{0}U_{0,\sigma\sigma\sigma}\right]
 			=-\alpha S_{0,\sigma\sigma}-U_{0,\tau}-\dfrac{1}{3}(S_{0})^{2}U_{0,\tau\sigma\sigma}+S_{0}S_{0,\sigma}U_{0,\tau\sigma}.
 	\end{align}}
 \end{subequations}\vspace{-6pt}
 
 The set of Equations \eqref{GN order 0} and \eqref{GN order 1} can be reformulated in a matrix. Indeed, Equation \eqref{GN order 0} can be written equivalently as follows:
 \begin{equation}\label{ordre 0}
 	\left( {\begin{array}{cc}
 			\widehat{A}_{0} & \widehat{B}_{0} \\
 			\widehat{C}_{0} & \widehat{D}_{0} \\
 	\end{array} } \right)
 	\left( {\begin{array}{cc}
 			U_{0}	 \\
 			S_{0} \\
 	\end{array} } \right)=
 	\left( {\begin{array}{cc}
 			0  \\
 			0  \\
 	\end{array} } \right),
 \end{equation}
 where
 \begin{subequations}
 	\begin{align}
 		&\widehat{A}_{0}=S_{0}\partial_{\sigma},\\
 		&\widehat{B}_{0}=-v\partial_{\sigma}+U_{0}\partial_{\sigma},\\
 		&\widehat{C}_{0}=-v\partial_{\sigma}+U_{0}\partial_{\sigma}
 		+\dfrac{1}{3}(S_{0})^{2}\left[v\partial_{\sigma\sigma\sigma}-U_{0}\partial_{\sigma\sigma\sigma}+U_{0,\sigma}\partial_{\sigma\sigma}\right],\\
 		&\widehat{D}_{0}=g\partial_{\sigma}+\big[vU_{0,\sigma\sigma}-U_{0}U_{0,\sigma\sigma}+(U_{0,\sigma})^{2}\big]S_{0}\partial_{\sigma}.
 	\end{align}
 \end{subequations}
 Similarly, Equation \eqref{GN order 1} can be expressed as follows:
 \begin{equation}\label{ordre 1}
 	\left( {\begin{array}{cc}
 			\widehat{A}_{1} & \widehat{B}_{1} \\
 			\widehat{C}_{1} & \widehat{D}_{1} \\
 	\end{array} } \right)
 	\left( {\begin{array}{cc}
 			U_{1} \\
 			S_{1} \\
 	\end{array} } \right)=
 	\left( {\begin{array}{cc}
 			E_{1}  \\
 			E_{2} \\
 	\end{array} } \right),
 \end{equation}
 where
 \begin{subequations}
 	\begin{align}
 		\widehat{A}_{1}=&S_{0}\partial_{\sigma}+S_{0, \sigma},\\
 		\widehat{B}_{1}=&-v\partial_{\sigma}+U_{0}\partial_{\sigma}+U_{0, \sigma},\\
 		\widehat{C}_{1}=&-v\partial_{\sigma}+U_{0}\partial_{\sigma}+U_{0, \sigma}+	S_{0}S_{0,\sigma}\left[v\partial_{\sigma\sigma}-U_{0}\partial_{\sigma\sigma}-U_{0,\sigma\sigma}+2U_{0,\sigma}\partial_{\sigma}\right]
 		\nonumber \\
 		&+\dfrac{1}{3}(S_{0})^{2}\left[v\partial_{\sigma\sigma\sigma}-U_{0}\partial_{\sigma\sigma\sigma}-U_{0,\sigma\sigma\sigma}+U_{0,\sigma}\partial_{\sigma\sigma}+U_{0,\sigma\sigma}\partial_{\sigma}\right],
 		\\
 		\widehat{D}_{1}=&g\partial_{\sigma}+\big[vU_{0,\sigma\sigma}-U_{0}U_{0,\sigma\sigma}+(U_{0,\sigma})^{2}\big](S_{0}\partial_{\sigma}+S_{0,\sigma})\nonumber
 		\\
 		&+\dfrac{2}{3}S_{0}\left[vU_{0,\sigma\sigma\sigma}-U_{0}U_{0,\sigma\sigma\sigma}+U_{0,\sigma}U_{0,\sigma\sigma}\right],\\
 		E_{1}=& -S_{0,\tau}        ,\\
 		E_{2}=&     -\alpha S_{0,\sigma\sigma}-U_{0,\tau}+\dfrac{1}{3}(S_{0})^{2}U_{0,\tau\sigma\sigma}+S_{0}S_{0,\sigma}U_{0,\tau\sigma}    .
 	\end{align}
 \end{subequations}
 
 \subsection{Application of Green's Theorem in One Dimension}
 
 Green's theorem in one dimension has been proved and applied to linear differential operators \cite{bams/1183416937,doi:10.1142/5740,doi:10.1142/0752}. Here, by extension, we applied this theorem to matrix differential operators. To do so, we briefly recalled the following theorem:
 \begin{equation}\label{Green 1 dim}
 	\int_{a}^{b}\left[  zP(y)-y\bar{P}(z)\right]dx=\left[  P(y,z)\right]_{a}^{b},
 \end{equation}
 where $ P $ is a linear differential operator and $ y $ and $ z $, any two functions of $ x $ and, $ \bar{P}(z) $ and $ P(y,z) $, are the adjoint and the bilinear differential expressions of $ P(y) $, respectively \cite{DarbouxLeonsSL}. This theorem, in its usual form, as it is shown in \eqref{Green 1 dim}, was previously used to show the damping of solitary waves \cite{doi:10.1063/1.1693097,doi:10.1063/1.1692358}. 
 
 In our case, we considered operators $\widehat{L}_{0}  $ and $ \widehat{L}_{1} $, defined as follows:
 \begin{center}
 	$ \widehat{L}_{0}= 	\left( {\begin{array}{cc}
 			\widehat{A}_{0} & \widehat{B}_{0} \\
 			\widehat{C}_{0} & \widehat{D}_{0} \\
 	\end{array} } \right), \qquad \widehat{L}_{1}=\left( {\begin{array}{cc}
 			\widehat{A}_{1} & \widehat{B}_{1} \\
 			\widehat{C}_{1} & \widehat{D}_{1} \\
 	\end{array} } \right). $ 
 \end{center}
 
 Using $  \widehat{L}_{0}  $ and  $\widehat{L}_{1} $, Equations \eqref{ordre 0} and \eqref{ordre 1} become
 \begin{equation}\label{V0 V1}
 	\widehat{L}_{0}V_{0}=0 \quad\text{and} \quad \widehat{L}_{1}V_{1}=E,
 \end{equation}
 where
 \begin{center}
 	$  V_{0}= \left( {\begin{array}{cc}
 			U_{0}	 \\
 			S_{0} \\
 	\end{array} } \right),\qquad   V_{1}= \left( {\begin{array}{cc}
 			U_{1}	 \\
 			S_{1} \\
 	\end{array} } \right), \qquad E=\left( {\begin{array}{cc}
 			E_{1}  \\
 			E_{2} \\
 	\end{array} } \right). $ 
 \end{center}
 
 Taking into account the symmetric behaviour of  $ S(x,t) $ and $ U(x;t) $ at $  x=\pm\infty $, the extension of Green's theorem in one dimension to linear differential matrix operator $\widehat{L}_{1}$~yields
 \begin{equation}\label{Green}
 	\int_{-\infty}^{+\infty}\left(V^{\dagger}_{0}\widehat{L}_{1}V_{1}
 	-V^{\dagger}_{1}\widehat{L}_{1}V_{0}\right)
 	d\sigma=0,
 \end{equation}
 where $ V^{\dagger}_{0} $ and $ V^{\dagger}_{1} $ are $ V_{0} $  and   $ V_{1} $ transposed, respectively. This extension can easily be proved following the procedure proposed in the original work of \cite{bams/1183416937}.
 
 \subsection{Blow-Up in Finite Time of the Serre-Green-Naghdi Soliton Solution}
 
 Using Equations \eqref{V0 V1} and \eqref{Green}, we have:
 \begin{equation}\label{integral 1}
 	\int_{-\infty}^{+\infty}\bigg[-U_{0}S_{0,\tau}+S_{0}\left(
 	-\alpha S_{0,\sigma\sigma}-U_{0,\tau}+\dfrac{1}{3}(S_{0})^{2}U_{0,\tau\sigma\sigma}+S_{0}S_{0,\sigma}U_{0,\tau\sigma}\right)
 	\bigg]d\sigma=0,
 \end{equation}
 which can be written as follows:
 \begin{equation}\label{integral 2}
 	-	\int_{-\infty}^{+\infty}\bigg[\dfrac{\partial}{\partial \tau}\left(U_{0}S_{0,}\right)
 	\bigg]d\sigma 
 	+	\alpha\int_{-\infty}^{+\infty}(S_{0,\sigma})^{2}d\sigma
 	-\alpha\bigg[S_{0}S_{0,\sigma}	\bigg]_{-\infty}^{+\infty}
 	+\bigg[\dfrac{1}{3}(S_{0})^{3}U_{0,\tau\sigma}
 	\bigg]_{-\infty}^{+\infty}=0,
 \end{equation}
 where $ U_{0} $ and $ S_{0} $ are the unperturbed solutions with a time-dependent amplitude, namely,
 \begin{align}
 	U_{0}=&c_{0}\left(
 	1+\frac{a(\tau)}{h}
 	\right)^{1/2}\left(
 	1-\frac{h}{S_{0}}
 	\right), \label{U0 solution}\\
 	S_{0}=&\frac{a(\tau)}{[\cosh(\beta)]^{2}}, \qquad 
 	\beta=\sqrt{\frac{3}{4}}\left(\frac{1}{h}\right)\left(\frac{a(\tau)}{a(\tau)+h}\right)^{1/2}
 	\left[
 	x-c_{0}t\left(
 	1+	\frac{a(\tau)}{h}
 	\right)^{1/2}
 	\right].\label{S0 solution}
 \end{align}
 
 After inserting \eqref{S0 solution} and \eqref{U0 solution} into Equation \eqref{integral 2}, alongside Equation \eqref{change of variable} it can be noticed that the limit of $\displaystyle{\frac{1}{\cosh(\sigma)}} $ tends to zero, while $ \sigma\to \pm \infty $. The Equation \eqref{integral 2} yields the following:
 \begin{equation}\label{integral 3}
 	\int_{a(0)}^{a(\tau)}\frac{(a+h)^{1/2}(2a+h)}{a^{3}}da=\frac{4\alpha}{5c_{0}}\left(\frac{1}{h}\right)^{3/2}\tau.
 \end{equation}

 Using the variable $ r $, Equation \eqref{integral 3} yields:
 \begin{equation}\label{integral 4}
 	\int_{r(0)}^{r(\tau)}\frac{(1+r)^{1/2}(1+2r)}{r^{3}}dr=
 	\frac{4\alpha}{5c_{0}h}\tau.
 \end{equation}
 
 The series expansion for a small value of $ r $ in the Equation yields:
 \begin{equation}\label{integral 5}
 	\int_{r(0)}^{r(\tau)}
 	\left(\frac{1}{r^{3}}+\frac{5}{2r^{2}}+\frac{7}{8r}-\frac{3}{16}+o(r)
 	\right)
 	dr=
 	\frac{4\alpha}{5c_{0}h}\tau.
 \end{equation}
 
 After integration, Equation \eqref{integral 5} becomes
 \begin{equation}\label{integral 6}	
 	\left[
 	-\frac{3r}{16}+\frac{7\ln(r)}{8}-\frac{5}{2r}-\frac{1}{2r^{2}}+o(r^{2})
 	\right]_{r(0)}^{r(\tau)}=
 	\frac{4\alpha}{5c_{0}h}\tau,
 \end{equation}
 
 and by keeping the leading term for $ r $ small, Equation \eqref{integral 6} becomes
 \begin{equation}\label{integral 7}	
 	\left[
 	-\frac{1}{2r^{2}}
 	\right]_{r(0)}^{r(\tau)}=
 	\frac{4\alpha}{5c_{0}h}\tau.
 \end{equation}
 
 Hence, the time dependent amplitude reads as follows:
 \begin{equation}\label{t-amplitude}
 	a(\tau)=\frac{A_{0}}{\sqrt{1-\displaystyle{\frac{8\alpha A^{2}_{0}\tau}{5c_{0}h^{3}}}}},
 \end{equation}
 where $ A_{0}=a(0) $. From Eq. \eqref{t-amplitude}, it can be seen that the amplitude $ a(\tau) $ tends to infinity, when the slow time $ \tau $ approaches a certain value $ \tau_{b} $, which we call the \textit{``slow'' blow-up time}. Replacing $ \alpha=\epsilon \Delta^{2} $, and  $ \tau=st $, the  blow-up time can be written as follows:
 \begin{equation}\label{t_b}
 	t_{b}=\frac{5c_{0}h^{3}}{8\epsilon \Delta^{2}A^{2}_{0}s},
 \end{equation}
 where  $\Delta^{2}=\left(U_{10}-C_{GN}\right)^{2} $, and the dispersion relation of the SGN solitary wave $ C_{GN} $, can be written \cite{GN2, Manna_2018} as follows:
 \begin{equation}\label{GN dispersion relation}
 	C_{GN}=\frac{c_0}{\sqrt{1+\frac{1}{3}(kh)^{2}}}.
 \end{equation}
 
 \subsection{Blow-up Time's Effective evaluation }\label{sec3}
 %%%%%%%%%%%%%%%%%%%%%%%%%%%%%%%%%%%%%%%%%%%%%%%%%%%%%%%%%%%%%%%%%%%%%%%%%%%%%%%%%%%%
 In order to effectively evaluate the blow-up time $ t_b $ and the growth rate of 
 wind waves in finite depth, we used detailed measurements of shallow water parameters in finite depth experiments conducted in the IRPH\'E/Pyth\'eas wind-wave tank \cite{BRANGER2022104174}. These measurements were carried out for non-dimensional depth $kh$ and non-dimensional initial waves' peak value $ kA_0 $. This led us to consider the non-dimensional soliton-like solution (a solution which at any time looks exactly as a soliton but with parameters which are triggered by the interaction with the source) $ kS_{0} $, instead of $ S_0 $, as well as the non-dimensional amplitude $ ka$ that we denoted as a function of non dimensional time $ \overline{t} $, as follows:
 \begin{equation}\label{non dimentional amplitude}
 	ka(\overline{t})=\frac{kA_{0}}{\sqrt{1-\displaystyle{\frac{8\epsilon  (kA_{0})^{2}s}{5(kh)^{2}}}\hskip3pt\overline{t}}},
 \end{equation}
 where
 \begin{equation}\label{non dimensional time}
 	\overline{t}=\frac{\Delta^{2}}{c_{0}h}t.
 \end{equation}
 
 Notice that the values of $ c_0 $ as well as  $ U_{10} $ were also measured  experimentally. For this reason, in what follows, the values of $ c_0 $ differ slightly from the theoretical values $c_0=\sqrt{gh}  $. 
 
 Using the experimental data of IRPH\'E/Pyth\'eas facilities (\cite{BRANGER2022104174}), for
 \begin{align}\label{experimental data}
 	h=0.14 \hskip3pt \text{m}, \quad	kh=1.54, \quad kA_0=0.114, \quad   c_0=0.92\hskip3pt \text{m/s}, \quad U_{10}=4.82\hskip3pt \text{m/s}, 
 \end{align}
 for the sheltering coefficient, we used $\epsilon=0.5 $, and for the small parameter we used $ s=0.001 $. 
 Using experimental data \eqref{experimental data} and Equations \eqref{t_b} and \eqref{GN dispersion relation} assisted in calculating the value of the blow-up time:
 \begin{equation}
 	t_b\approx 1721 \hskip4pt \text{s}.
 \end{equation}

 The $ x $-position of the SGN soliton-like solution as a function of time is found using
 \begin{equation}\label{x(t)}
 	x(t)=c_{0}t\left(
 	1+	\frac{a(\tau)}{h}
 	\right)^{1/2}.
 \end{equation}
 
 The length of the IRPH\'E/Pyth\'eas wind--wave tank facility was 40 m. The growing solitary wave reaches the tank's end after 40--45 s. Consequently,
 the wave amplitude is not nearly that of the blow-up. The growth rate of the solitary wave at different times and positions, for $ h=0.14\hskip3pt \text{m} $, and $ U_{10}=4.48\hskip3pt \text{m/s} $, is given in (Table \ref{tab1}). In these conditions, when the wave reaches the tunnel's end, the growth rate is approximately $ 0.1 $. Hence, it is at the measurability limit of the IRPH\'E/Pyth\'eas wind--wave tank facility.
 
 \begin{table}
 	\caption{The growth rate of the SNG solitary wave at different times and positions, for depth $ h=0.14\hskip3pt \text{m} $, and wind speed of 10 m $ U_{10}=4.82\hskip3pt \text{m/s}. $\label{tab1}}
 	\newcolumntype{C}{>{\centering\arraybackslash}X}
 	\begin{tabularx}{\textwidth}{CcCCCCCC}
 		\toprule
 		\textbf{{\emph{t} (s)}} %MDPI: Is the bold format in this table necessary? Should this column be the tale header? If yes, please make it the first row of the table. The same problem in Table 2.
 		& 0	& 40 & ...	& 750 &  1000& 1250	& 1400	\\	
 		\midrule
 		\textbf{{\emph{x} (m)}}& 0	& 35
 		& ... & 722 & 972	& 1229 & 1329
 		\\	
 		\midrule
 		\textbf{{growth~rate}}&	& 0.09	& ... &0.15 &0.55 & 1	& 1.36	\\
 		\bottomrule
 	\end{tabularx}
 \end{table}
 \unskip
 %This result is in agreement with the wave velocity $ c_0=0.95 $ m/s. 
 And for
 \begin{align}\label{experimental data 2}
 	h=0.26 \hskip3pt \text{m}, \quad	kh=2.57,   \quad kA_0=0.146,  \quad 
 	c_0=1.0 \hskip3pt \text{m/s}, \quad  U_{10}=4.35\hskip3pt \text{m/s},
 \end{align}
 with a sheltering coefficient $\epsilon=0.5 $ and small parameter $ s=0.001 $. 
 Similarly, experimental data \eqref{experimental data 2} and Equations \eqref{t_b} and  \eqref{GN dispersion relation} led to the corresponding blow-up time:
 \begin{equation}
 	t_b\approx 7008 \hskip4pt \text{s}.
 \end{equation}

 The continuous growth of ($ kS_0 $), leads to blow-up at finite time, $ t_b\approx$ 7008 s, which of course, is out of reach. Therefore, in this case, a significant growth in the solitary wave's amplitude was not observable in experimental facilities, but it could be \textit{in situ}. The growth rate of the solitary wave solution at different times and positions, for $ h= 0.26 $  m, and $ U_{10}=4.35\hskip3pt \text{m/s} $, is given in (Table \ref{tab2}).
 
 \begin{table}
 	\caption{The growth rate of the SNG solitary wave at different times and positions, for depth $ h=0.26\hskip3pt \text{m} $, and wind speed at 10 m $ U_{10}=4.35\hskip3pt \text{m/s}. $\label{tab2}}
 	\newcolumntype{C}{>{\centering\arraybackslash}X}
 	\begin{tabularx}{\textwidth}{CcCCCCCC}
 		\toprule
 		
 		\textbf{{\textit{t} (s)}} & 0	& 40 & ...	& 750 &  1000 & 1250 & 15,000	\\	
 		\midrule
 		\textbf{{\textit{x} (m)}}& 0	& 41	 & ... & 772 & 1030	& 1288 & 1547
 		\\	
 		\midrule
 		\textbf{{growth~rate}}&	& 0.02	& ... &0.06 &0.08 & 0.1 & 0.12	\\
 		\bottomrule
 	\end{tabularx}
 \end{table}
 \unskip

 %%%%%%%%%%%%%%%%%%%%%%%%%%%%%%%%%%%%%%%%%%%%%%%%%%%%%%%%%%%%%%%%%%%%%%%%%%%%%%%%%%
 
 %%%%%%%%%%%%%%%%%%%%%%%%%%%%%%%%%%%%%%%%%%%%%%%%%%%%%%%%%%%%%%%%%%%%%%%%%%%%%%%%%%%%%%%%%%%%%

 %%%%%%%%%%%%%%%%%%%%%%%%%%%%%%%%%%%%%%%%%%%%%%%%%%%%%%%%%%%%%%%%%%%%%%%%%%%%%%%%%%%%%%%%%
 %%%%%%%%%%%%%%%%%%%%%%%%%%%%%%%%%%%%%%%%%%%%%%%%%%%%%%%%%%%%%%%%%%%%%%%%

\section{Korteweg-de Vries-Burger equation}

The Jeffreys' theory allows to compute the linear wave 
growth of wind-generated normal Fourier modes of wave-number $k$. The physical mechanism 
	behind is "focusing", in the sense that energy passes continuously from the air to the surface wave.
Consequently the wave amplitude $\eta(x,t,k)$ ($x$ space and $t$ time) grows 
exponentially in time i.e.; $\eta(x,t,k)\sim\exp{(\gamma_{J} t)}$ more or less quickly according to the coefficient $\gamma_{J}$, which depends on the wind speed and the water depth $h$.
Once the linear and dispersionless approximation breaks down, non-linear and dispersive processes begin to play a role.
So the issue addressed here is: "how to describe the evolution in time of a normal mode $k$, under the competing actions of (weak) nonlinearity,
	dispersion and anti-dissipation in the sense of a contiuous energy transfer from wind to water?" 
Nonlinearity is likely to balance dispersive effects, or to stop exponential decay or growth
of wave amplitude in time due to dissipation or "anti-dissipation". Equilibrium between nonlinearity and dispersion
can evolve in time to form solitary waves as in the Korteweg-de Vries equation \cite{Whitham,KdV}. Balance between dissipation or "anti-dissipation" and nonlinearity creates shock 
structures as in the Burgers equation \cite{Whitham}. The standard equation describing competition between weak non-linearity, dispersion and dissipation is the
KdV-B equation.
It appears in many physical contexts 
\cite{Benney, Johnson, Grad1,Grad2, Wadati, Kara}. In this section, in order to study simultaneous competing effects of weakly nonlinearity, dispersion and
"anti-dissipation" we derive a KdV-B type equation
with {\it dissipation turned into "anti-dissipation"}.\\

Green-Naghdi equations under the wind action \cite{GN1, GN2,Manna_2018} are given by Eqs \eqref{GNW1} and \eqref{GNW2}. We introduce dimensionless "primed" variables, $ x' $, $ t' $ and $ \eta' $ as follows:
\begin{equation}\label{primed variables}
	x=\lambda_0x', \qquad t= \frac{\lambda_0}{c_0}t' , \qquad  \eta=a_0\eta' ,
\end{equation}
where $ c_0=(gh)^{1/2} $, $ a_0 $ and $ \lambda_0 $ are initial typical wave amplitude and wavelength, respectively.  In addition, we define two fundamental parameters, commonly used in the classical water surface studies, namely, $ \nu $ and $ \delta $, as follows:
\begin{equation}\label{nu-lambda}
	\nu=\frac{a_0}{h}, \qquad \delta=\frac{h}{\lambda_0}.
\end{equation}
Finally, in order to obtain the dimensionless, scaled Green-Naghdi equations of motion, the following scaling is required \cite{Johnson}:
\begin{equation}\label{J-scaling}
	u=\nu u_0.
\end{equation}
Introducing Eqs \eqref{primed variables}, \eqref{nu-lambda} and \eqref{J-scaling} in Eqs \eqref{GNW2} and \eqref{GNW1}, we obtain the following dimensionless equations
\begin{subequations}\label{J-GNWind}
	\begin{align}
		& u_{0,t}+\nu u_0u_{0,x}+\eta_x   =-\epsilon s \Delta^{2}\delta\eta_{xx}+\frac{\delta^{2}}{3(1+\nu\eta)}\{(1+\nu\eta)^3(u_{0,xt}+\nu u_0u_{0,xx}-\nu u^2_{0,x})\}_x,\label{J-GNW2} \\
		&\eta_t+ [u_0(1+\nu\eta ) ]_x=0, \label{J-GNW1}
	\end{align}
\end{subequations}
where, for convenience, the "primes" of dimensionless quantities are omitted. Now, we consider a wave moving from left to right  \cite{Whitham}. At the lowest order, by neglecting the terms of order    $ \nu $ and $ \delta $ and any higher orders, Eqs \eqref{J-GNW2} and \eqref{J-GNW1} are reduced to
\begin{subequations}
	\begin{equation}\label{reduction 1}
		u_{0,t}+\eta_{x}=0,
	\end{equation}
	\begin{equation}\label{reduction 2}
		\eta_{t}+u_{0,x}=0.
	\end{equation}
\end{subequations}
Eqs \eqref{reduction 1} and \eqref{reduction 2} are equivalent to
\begin{equation}
	\eta_t+\eta_x=u_{0,t}+u_{0,x}=0,
\end{equation}
and its solution is
\begin{equation}
	u_{0,x}(x,t)=\eta(x,t).
\end{equation}
Now we look for a perturbed solution with follow form:
\begin{equation}\label{perturbation 1}
	u_0=\eta+\nu	\mathbb{A}+\delta \mathbb{B}+\delta^{2}\mathbb{C}+O(\delta\nu,\delta^{2}\nu,\nu^{2}),
\end{equation}
where $ \mathbb{A} $, $ \mathbb{B} $ and $ \mathbb{C  } $ are functions of $ \eta $ and its derivatives. Inserting Eq. \eqref{perturbation 1} in \eqref{J-GNW2} and \eqref{J-GNW1}, we obtain
\begin{subequations}
	\begin{align}
		&	\eta_t+\eta_{x}+\nu(\mathbb{A}_t+\eta\eta_x)+\delta(\mathbb{B}_t+\epsilon s\Delta^{2}\eta_{xx})+\delta^{2}(\mathbb{C}_t-\frac{1}{3}\eta_{xxt})+O(\delta\nu,\delta^{2}\nu,\nu^{2})=0, \label{first1} \\
		&	\eta_t+\eta_{x}+\nu(\mathbb{A}_x+\eta\eta_x)+\delta(\mathbb{B}_x+\delta^{2}\mathbb{C}_x+O(\delta\nu,\delta^{2}\nu,\nu^{2})=0,\label{second1}
	\end{align}
\end{subequations}
where
\begin{equation}
	\eta_{t}=-\eta_{x}+O(\delta\nu,\delta^{2}\nu,\nu^{2}).
\end{equation}
Therefore, in Eq. \eqref{first1}, all the $ t $-derivatives may be substituted by $ -\partial_x $. Hence, Eqs \eqref{first1} and \eqref{second1} are compatible, if
\begin{equation}\label{ABC}
	\mathbb{A}=-\frac{1}{4}\eta^{2}, \qquad \mathbb{B}=\frac{1}{2}s\epsilon\Delta^{2}\eta_{x}, \qquad \mathbb{C}=\frac{1}{6}\eta_{xx}.
\end{equation}
Substituting Eq. \eqref{ABC} in Eqs \eqref{first1} and \eqref{second1}, yields
\begin{subequations}
	\begin{align}
		&		\eta_t+\eta_x+\frac{3}{2}\nu\eta\eta_x+\frac{\delta^{2}}{6}\eta_{xxx}+s\epsilon\frac{\delta}{2}\Delta^{2}\eta_{xx}=0,\label{kdvb}\\
		&u-\eta+\frac{\nu}{4}\eta^{2}-s\epsilon\frac{\delta}{2}\Delta^{2}\eta_{x}-\frac{1}{6}\delta^{2}\eta_{xx}=0.\label{invarant}
	\end{align}
\end{subequations}
Eq. \eqref{kdvb} is the KdV-B equation, while Eq. \eqref{invarant} is a Riemann invariant.

\subsection{Solution of  Korteweg-de Vries-Burger equation}

In order to find the solution of Eq.\eqref{kdvb}, we apply the following change of variables:
\begin{equation}
	\sigma=x-t, \qquad t_1=\delta^{2}t.
\end{equation}
Hence, Eq.\eqref{kdvb} becomes:
\begin{equation}\label{KdV-B2}
	\eta_{t_1}+\frac{3}{2}\eta\eta_{\sigma}+\frac{\delta^{2}}{6\nu}\Delta^{2}\eta_{\sigma\sigma\sigma}+\frac{\delta}{2\nu}\Delta^{2}s\eta_{\sigma\sigma}=0.
\end{equation}
It worth noticing that the limit of Eq. \eqref{KdV-B2}, as $ s\to 0 $, yields to the well known KdV equation:
\begin{equation}\label{kdv}
	\eta_{t_1}+\frac{3}{2}\eta\eta_{\sigma}+\frac{\delta^{2}}{6\nu}\Delta^{2}\eta_{\sigma\sigma\sigma}=0.
\end{equation}
This result is quite natural, since  $ s\to 0 $, amounts to neglecting the action of the wind. Therefore, it is possible to assume that the solution  KdV-B, Eq. \eqref{KdV-B2}, has the same form as  the  solution of Eq.   \eqref{kdv}, namely
\begin{equation}\label{kdv solution}
	\eta(\sigma, t_1)=\frac{a}{\cosh^{2}\left[P(\sigma-c't_1)\right]}.
\end{equation}
This is the typical "soliton" solution of KdV equation, with one difference however, the amplitude $a $ in Eq. \eqref{kdv solution} can be time-dependant, whereas the amplitude of KdV equation is not. 

Inserting \eqref{kdv solution} in \eqref{KdV-B2}, we obtain
\begin{equation}
	P=\sqrt{\frac{3\nu}{4\delta^{2}}}, \qquad c'=\frac{a}{2}.
\end{equation}

Now, the task is to find the time-dependent expression of $ a(t) $ in Eq. \eqref{kdv solution}.

Noticing that the anti diffusive term $\displaystyle{\frac{\delta}{2\nu}\Delta^{2}s\eta_{\sigma\sigma}} $  in Eq. \eqref{KdV-B2}, is of order $ \delta^{3} $ and small enough  at $ t=0 $, one can find the solution of Eq. \eqref{KdV-B2} by perturbation. For this purpose,  we introduce a slow time $ t_2 $, as follows
\begin{equation}\label{slow time t_2}
	t_2=\delta^{2}t_1=\delta^{3}t,
\end{equation}
and we expand $ \eta $ in terms of $ \delta $ as follows
\begin{equation}\label{perturbation}
	\eta=\eta_{0}(\sigma ,t_2)+\delta \eta_1 (\sigma ,t_2)+O(\delta^{2}),
\end{equation}
where $ \eta_0 $ is the solution given by Eq. \eqref{kdv solution}. 

Introducing $ \nu_0  $ as follows
\begin{equation}
	\nu=\nu_0 \delta^{2}, 
\end{equation}
and inserting Eq. \eqref{perturbation} in Eq. \eqref{KdV-B2}, we obtain\\
\begin{subequations}\\
	\textit{at order 0 of $ \delta $:} 
	\begin{equation}\label{order0}
		\frac{\partial \eta_0}{\partial t_1}+\frac{3}{4}\nu_0\eta_0	\frac{\partial \eta_0}{\partial \sigma}+\frac{1}{6}\frac{\partial^{3} \eta_0}{\partial \sigma^{3}}=0,
	\end{equation}
	and, \textit{at order 1 of $ \delta $:}
	\begin{equation}\label{order1}
		\frac{\partial \eta_1}{\partial t_1}+\frac{3}{2}\nu_0\eta_0	\frac{\partial \eta_1}{\partial \sigma}+\frac{3}{2}\nu_0\eta_{0,\sigma}\eta_1+\frac{1}{4}\nu_0	\frac{\partial^{3} \eta_1}{\partial \sigma^{3}}=-\eta_{0,t_2}-\frac{s_0}{2}\epsilon\Delta^{2}_{0}\eta_{0,\sigma\sigma}.
	\end{equation}
\end{subequations}
Introducing operators $ \widehat{L}_0 $ and $ \widehat{L}_1 $ as follows
\begin{subequations}\\ 
	\begin{equation}\label{L_0}
		\widehat{L}_0=	\frac{\partial }{\partial t_1}+\frac{3}{4}\nu_0\eta_0	\frac{\partial }{\partial \sigma}+\frac{1}{6}\frac{\partial^{3} }{\partial \sigma^{3}},
	\end{equation}
	\begin{equation}\label{L_1}
		\widehat{L}_1=	\frac{\partial }{\partial t_1}+\frac{3}{2}\nu_0\left(\eta_0	\frac{\partial }{\partial \sigma}+\eta_{0,\sigma}\right)+\frac{1}{4}\nu_0	\frac{\partial^{3} }{\partial \sigma^{3}}.
	\end{equation}
\end{subequations}
Eqs \eqref{L_0} and \eqref{L_1} read
\begin{subequations}\\ 
	\begin{equation}\label{first}
		\widehat{L}_0\eta_0=0,
	\end{equation}
	\begin{equation}\label{second}
		\widehat{L}_1\eta_1=	-\eta_{0,t_2}-\frac{s_0}{2}\epsilon\Delta^{2}_{0}\eta_{0,\sigma\sigma}.
	\end{equation}
\end{subequations}

To go further, we apply  Green's theorem in one dimension. One can find the application of this theorem  to linear differential operators in various works \cite{bams/1183416937,doi:10.1142/5740,doi:10.1142/0752}. In particular, the  damping of solitary waves \cite{doi:10.1063/1.1693097,doi:10.1063/1.1692358} has been shown using this theorem and an extension to matrix differential operators has been performed by \cite{fluids7080266}. 

Applying Green's theorem to our case by replacing  Eqs \eqref{L_0} and \eqref{L_1}  in Eq. \eqref{Green 1 dim}, yields
\begin{equation}\label{our case}
	\int_{-\infty}^{+\infty}\left(  \eta_0 \widehat{L}_1\eta_1-\eta_1\widehat{L}_0\eta_0 \right)d\sigma=0.
\end{equation}
Notice that the right hand side of \eqref{our case} is null due to the symmetric behaviour of $ \eta_0 $ and $ \eta_1 $ at $ \pm \infty $. Replacing $ \eta_0 $ by Eq. \eqref{kdv solution} with a time dependant amplitude $ a(t_2) $,  Eq. \eqref{our case} yields:
\begin{equation}\label{blow up t_2}
	a(t_2)=\frac{1}{1-\frac{2}{5}\frac{s\epsilon\Delta^{2}}{\delta^{2}}t_2}.
\end{equation}
Using the approximation $ O(\nu) =O(\delta^{2})$, Eq. \eqref{slow time t_2} can equivalently be written as $ t_2=\nu\delta t $. Hence, Eq. \eqref{blow up t_2} becomes
\begin{equation}\label{amplitude at}
	a(t)=\frac{1}{1-\frac{t}{t_b}},
\end{equation}
where 
\begin{equation}\label{blow up t}
	t_b=\frac{5\delta}{2\epsilon s\Delta^{2}\nu}.
\end{equation}
From Eq. \eqref{amplitude at}, it can be seen that the amplitude $ a(t) \to \infty $, when $ t \to t_b $ which we call the blow-up time.
Hence, the solution of \eqref{kdvb} reads:
\begin{equation}\label{kdvb soliton}
	\eta=a(t)\cosh^{-2}(\theta),
\end{equation}
where
\begin{equation}\label{details}
	\theta (x,t)=\alpha a^{1/2}\left[
	x-t+\frac{\nu}{2}t_b\ln\left(1-\frac{t}{t_b}\right)
	\right], \qquad 
	\alpha=\left(
	\frac{3}{4}\frac{a_0}{h}
	\right)^{1/2}\frac{\lambda_0}{h},
\end{equation}

\subsection{Blow-up  in finite time and the evolution of the solitary wave solution's shape}
In this section we shall study the evolution of the soliton-like solution's shape in time , Eq. \eqref{kdvb soliton},  before the blow up time $ t_b $.

Coming back to variables with dimensions, Eq. \eqref{primed variables}, the solution of KdV-B, i.e. Eq. \eqref{kdvb soliton} reads:
\begin{equation}\label{kdvb solition with dimensions}
	\eta(x,t)=\frac{a_0}{1-\frac{t}{t_b}}\cosh^{-2} \biggl\{
	\frac{\alpha}{(1-\frac{t}{t_b})^{1/2}}\frac{1}{\lambda_0}
	\left[
	x-c_0t+\frac{\nu}{2}c_0t_b\ln \left( 1-\frac{t}{t_b}\right)
	\right]
	\biggr\}.
\end{equation}
For $ t=0 $, we have
\begin{equation}
	\eta(x,0)=a_{0}\cosh^{-2}\left(\frac{\alpha}{\lambda_0}x\right),
\end{equation}
where
\begin{equation}\label{used in the table}
	\alpha^{2}=\frac{3\nu}{4\delta^{2}}=\frac{3}{4}\frac{a_0{\lambda_0}^{2}}{h^{3}}.
\end{equation}
Using Eq. \eqref{used in the table}, the blow-up time, $ t_b $ can also be expressed as follows for further use:
\begin{equation}\label{blow up 2}
	t_b=\frac{5}{2}\frac{c_0h^{2}}{\epsilon s a_0 \Delta^{2}}.
\end{equation}

The wave number $ k $ and the frequency $ \omega $, for a monochromatic progressive wave with a phase $ \theta(x,t)=kx-\omega t $, are defined as follows:
\begin{equation}
	k=\frac{\partial \theta}{\partial x}, \qquad \omega=-\frac{\partial \theta}{\partial t}.
\end{equation}
These definitions can be generalized for $ k $ and $ \omega $ depending on $ x $ and $ t $:
\begin{equation}\label{x an t dependant}
	k(x,t)=\frac{\partial \theta}{\partial x}(x,t), \qquad \omega(x,t)=-\frac{\partial \theta}{\partial t}(x,t).
\end{equation}
Using Eqs \eqref{x an t dependant}, \eqref{details} and \eqref{amplitude at}, the wavelength $ \lambda(t) $ and the wave number $ k(t) $ of the soliton-like solution \eqref{kdvb soliton} are
\begin{align}
	&	\lambda(t)=\frac{\lambda_0}{\alpha }\left(1-\frac{t}{t_b}\right)
	=\left(\frac{4h}{3a_0}\right)^{1/2}h\left(1-\frac{t}{t_b}\right)^{1/2},	\label{lambda de t}\\
	& k(t)=\left(\frac{3a_0}{4h}\right)^{1/2}\frac{1}{h}\left(1-\frac{t}{t_b}\right)^{-1/2}=\alpha a^{1/2}(t).\label{k de t}
\end{align}

We define the effective wave number, noted $\widetilde{k}$,  as follows:
\begin{equation}\label{lambda effectif}
	\widetilde{k}=\frac{\alpha}{\lambda_0}=\left(\frac{3a_0}{4h}\right)^{1/2}\frac{1}{h},
\end{equation}
and  we define the associated effective wavelength, noted  $\widetilde{\lambda}$, defined as follows:
\begin{equation}\label{ k effectif}
	\widetilde{\lambda}=\frac{1}{\widetilde{k}}=\left(\frac{4h}{3a_0}\right)^{1/2}h.
\end{equation}
It worth noticing that for $ t \to t_b $,  Eqs  \eqref{lambda de t}, \eqref{k de t} and\eqref{amplitude at} give
\begin{equation}
	\lim_{t\to t_b} \lambda(t)=0, \qquad 	\lim_{t\to t_b} k(t)=\infty, \qquad 	\lim_{t\to t_b} a(t)=\infty,
\end{equation}
respectively. This is the first indication of  the narrowing of  the solitary wave solution's shape while its amplitude $ a(t) $ grows.

To go further, we are going to examine the speed of phase planes on either side of the  soliton-like solution's crest. Using Eq. \eqref{k de t}, it is useful to write  Eq. \eqref{details} as a function of $ k(t) $ as follows
\begin{equation}
	\theta(x,t)=k(t)(x-t)+\nu t_bk(t) \left[ \ln(x) -\ln\left(k(t)\right)\right].
\end{equation}
Using Eq. \eqref{k de t}, we have
\begin{equation}\label{k_t}
	\frac{\partial k(t)}{\partial t}= \frac{k^{3}(t) }{2\alpha^{2}t_b}.
\end{equation}
Eqs \eqref{x an t dependant} and \eqref{k_t} together, yields $ \omega $ as a function of $ k(t) $:
\begin{equation}
	\omega(x,t)=-\frac{k^{3}(t)}{2\alpha^{2}t_b}(x-t)+k(t)-\nu t_b\frac{k^{3}(t)}{2\alpha^{2}t_b} \left[\ln\left(\frac{\alpha}{k(t)}-1\right)\right].
\end{equation}
The dimensionless phase velocity $ c(x,t) $ of the soliton-like solution is
\begin{equation}
	c(x,t)=\frac{\omega(x,t)}{k(t)}=1+\frac{\nu}{2\alpha^{2}}k^{2}(t)-\frac{k(t)\theta(x,t)}{2\alpha^{2}t_b},
\end{equation}
which, by using Eq. \eqref{k de t}, can equivalently be written as a function of $ a(t) $ as follows
\begin{equation}\label{soliton celerity}
	c(x,t)=1+\frac{\nu}{2}a(t)-\frac{a^{1/2}(t)}{2\alpha t_b}\theta(x,t).
\end{equation}
The position of solitary wave's crest is found by solving 
\begin{equation}
	\theta(x,t)=0,
\end{equation}
at any time $ t $.
The dimensionless phase $ \theta $ is given by the first of the equations in Eq. \eqref{details}, while the phase velocity $ c $ is given by Eq. \eqref{soliton celerity}. Therefore, the position of the solitary wave's crest and the velocity of the solitary wave at it's crest are 
\begin{align}
	&	x_{crest}(t)=t+\frac{\nu t_b}{2}\ln[a(t)],\\
	& c(x_{crest},t) =1+\frac{\nu}{2}a(t),\label{c crest}
\end{align}
respectively.\\

Using Eqs \eqref{soliton celerity} and \eqref{details}, the phase speed becomes:
\begin{align}
	c&=c_{crest} -\frac{\alpha}{2}\frac{a(t)}{t_b}\left[
	(x-t)-\frac{\nu}{2}t_b \ln\left[a(t)\right]
	\right]
	\nonumber \\
	&=c_{crest} -\frac{\alpha}{2}\frac{a(t)}{t_b}\left[
	x-x_{crest}
	\right].
\end{align}
Therefore, the speed of the  phase planes at $( x_{crest}-\Delta x )$ and $( x_{crest}+\Delta x) $
are
\begin{subequations}
	\begin{equation}\label{right}
		c_{crest-\Delta x} =c_{crest}+\frac{\alpha}{2}\frac{a(t)}{t_b}\Delta x,
	\end{equation}
	\begin{equation}\label{left}
		c_{crest+\Delta x} =c_{crest}-\frac{\alpha}{2}\frac{a(t)}{t_b}\Delta x,
	\end{equation}
\end{subequations}
respectively. Hence, 
\begin{equation}
	c_{crest-\Delta x}>c_{crest}>	c_{crest+\Delta x}.
\end{equation}

This means that the phase planes at the left side of $ x_{crest} $ have greater speed than the phase planes at the right side of $ x_{crest} $ resulting to a narrowing the soliton-like solution's shape while it's amplitude $ a(t) $ grows (Figure 7).\\

At the first glance, the wave breaking does not result directly from the solitary wave solution of KdV-B equation. However, knowing that we have an \textit{accelerating} solitary wave at the speed $ c_{crest}$, the breaking might result from phase planes with non equal accelerations. This will be the subject of fore coming studies.

In the following subsections, various criteria of wave breaking are reviewed. 

\begin{figure}[h!]
	\centering
	\includegraphics[width=0.8\linewidth]{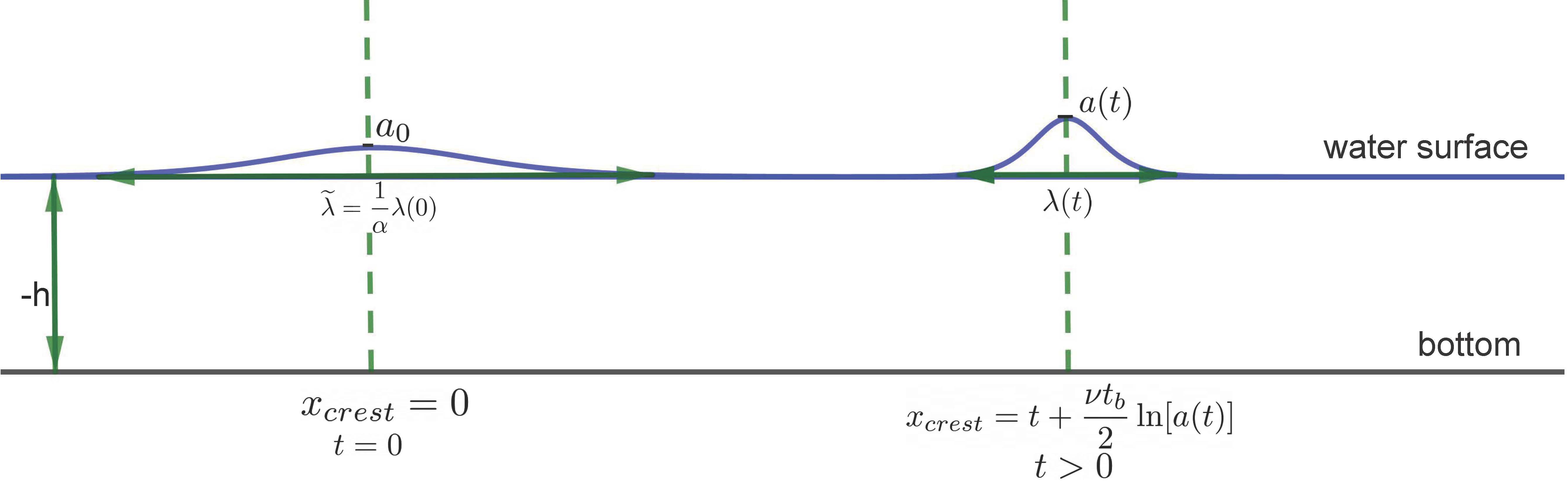}
	\caption{On the left, the soliton-like solution is plotted at $ t=0 $. The amplitude at crest is $ a_0 $ and its effective wavelength is $ \widetilde{\lambda} $. The origin of coordinates is placed at its crest position. On the left side, at $ t>0 $, the solitary wave has moved to the right, its shape has sharpened, $ \lambda(t)<\widetilde{\lambda} $, and its amplitude has grown, $ a(t)>a_0 $. Notice that for convenience, scales are not respected. }
	\label{fig:solitons}
\end{figure}

\subsection{Wave breaking criteria}

As noticed in the previous section, the wave breaking does not result from the evolution of the wave as the soliton-like solution of KdV-B equation. Therefore, we have to consider other reasons for the wave breaking. The most well-known criteria of wave breaking for  linear monochromatic waves are the McCowan criterium \cite{McCowan}, The Miche criterium \cite{Miche} and the horizontal velocity criterium \cite{Shemer}. But, as we shall see, this latest cannot be applied to our case. Therefore, we propose an alternative criterium that we shall call "alternative velocity criterium". For further use, we  call $ t_{d,Mc} $,  $ t_{d,Mi} $ and $ t_{d,alv} $, the wave breaking times within McCowan criterium, Miche criterium and alternative velocity criterium, respectively.

In what follows, we shall compute the breaking time $ t_d $, as well as the amplitude, the wavelength and the phase velocity of the solitary wave at $ t=t_d $ within each of the mentioned criteria.

\subsection{McCowan criterium}

The McCowan criterium is established for linear solitary waves and studies the highest 
maximum height that such waves might attain without breaking. McCowan has shown that the breaking occurs at a given rate between the maximum wave amplitude $ a_{max} $ and the water depth $ h $. Namely:
\begin{equation}\label{The McCowan criterium}
	\frac{a_{max}}{h}\approx 0.78.
\end{equation}
Applying this criterium to the soliton-like solution Eq. \eqref{kdvb solition with dimensions}, where $ {a(t)={a_0}({1-\frac{t}{t_b}}
})^{-1} $, Eq. \eqref{The McCowan criterium} yields
\begin{equation}\label{breaking 1}
	\left(\frac{a_0}{h}\right)\frac{1}{1-\frac{t_{d,Mc}}{t_b}}\approx0.78.
\end{equation}
Eq. \eqref{breaking 1} gives the breaking time according to  McCowan criterium, which is
\begin{equation}\label{t_d McCowan}
	t_{d,Mc} \approx t_b(1-1.28\nu).
\end{equation}
Using Eq. \eqref{kdvb solition with dimensions}, it is easy to calculate  the maximum height reached at $ t=t_{d,Mc} $. This is
\begin{equation}\label{a_max McCowan}
	\eta_{max}=a_{max}\approx\frac{a_{0}}{1.28\nu}\approx 0.78 h.
\end{equation}
In addition Eq. \eqref{lambda de t}, one obtains  the solitary wave's wavelength at the moment of  wave breaking, namely $ \lambda(t_{d,Mc}) $:
\begin{equation}
	\lambda(t_{d,Mc})\approx 1.28 \left( \frac{4}{3} a_{0}h\right)^{1/2}.
\end{equation}
\subsection{Miche Criterium}
The Miche criterium is based on empirical observations of linear waves' dispersion relations in a finite depth $ h $ with a wave length $ \lambda $. The Miche criterium fixes the maximum height $ a_{max} $ reached by the wave before the break down as follows
\begin{equation}\label{Miche}
	\left(\frac{a}{\lambda}\right)_{max}=\frac{1}{7}\tanh\left(\frac{2\pi h}{\lambda}\right).
\end{equation}
In our case, replacing $ a $ by $ {a(t)={a_0}({1-\frac{t}{t_b}}
})^{-1} $, and $ \lambda $ by  Eq. \eqref{lambda de t}, in the right side of Eq. \eqref{Miche}, we have
\begin{equation}\label{left side Miche}
	\left(\frac{a}{\lambda}\right)_{max}=\frac{\nu}{2}(3\nu)^{1/2}\left(
	1-\frac{t_{d,Mi}}{t_b}
	\right)^{3/2}.
\end{equation}
Now, replacing $ \lambda $ by  Eq. \eqref{lambda de t} in the left side of Eq. \eqref{Miche},  using $ \nu=a_{0}/h $, taking $ t=t_{d,Mi} $ and expanding the hyperbolic tangent up of $ \displaystyle{\left(\frac{a}{\lambda}\right)_{max}} $ up to order $ (3\nu)^{3/2} $, we obtain
\begin{equation}\label{right side Miche}
	\frac{1}{7}\tanh\left(\frac{2\pi h}{\lambda}\right)=
	\frac{1}{7}\frac{\pi(3\nu)^{1/2}}{(1-\frac{t_{d,Mi}}{t_b})^{1/2}}-\frac{1}{21}
	\frac{\pi^{3}(3\nu)^{3/2}}{(1-\frac{t_{d,Mi}}{t_b})^{3/2}}+O(\nu^{5/2}).
\end{equation}
Using Eqs \eqref{Miche},  \eqref{left side Miche} and \eqref{right side Miche}, and neglecting terms of higher order than  $ \nu $, we obtain
\begin{equation}\label{t_d Miche}
	t_{d,Mi}=t_b\left(1-\frac{7\nu}{2\pi}-\pi^{2}\nu\right).
\end{equation}
The maximum height reached by the wave at $ t=t_{d,Mi} $ is
\begin{equation}\label{a_max Miche}
	\eta_{max}=a_{max}\approx\frac{0.09a_{0}}{\nu}\approx 0.09 h.
\end{equation}
Using Eqs \eqref{left side Miche} and \eqref{t_d Miche}, the solitary wave's wavelength can be calculated at $ t=t_{d,Mi} $, which is
\begin{equation}
	\lambda_{max}\approx 4.7\times 10^{-3} \left(\frac{h}{a_0}\right)^{5/2}h.
\end{equation}

\subsection{Alternative horizontal velocity criterium}
The horizontal velocity criterium assumes that the wave breaking occurs when the group velocity of a water plane wave exceeds the speed of the phase plane at crest. In our case, this criterium cannot be applied because, for a solitary wave, at first approach, the group velocity does not mean much. Therefore, alternatively we replace this criterium by an alternative approach which assume that the wave breaking occurs when the fluid horizontal velocity exceeds the phase plane velocity at crest. At this moment, the matter starts to be ejected from the wave.

The phase plane velocity is given by Eq. \eqref{soliton celerity}. The phase velocity at crest,  i.e. for $ \theta(x,t)=0 $, is given be Eq. \eqref{c crest}. The horizontal water velocity is given by Eq. \eqref{invarant}. Using Eqs \eqref{perturbation} and \eqref{J-scaling}, as well as  \eqref{slow time t_2} or equivalently $ t_2=\nu\delta t $, the water velocity can be expanded as follows
\begin{equation}
	u_0=\eta-\frac{\nu}{4}\eta^{2} +O(\delta,\delta^{2}).
\end{equation}
It can be seen from Eq.  \eqref{kdvb soliton} that at crest $ \eta(\theta=0)=a(t) $. Hence the alternative velocity criterium reads
\begin{equation}
	1+\frac{\nu}{2}a\geq	a-\frac{\nu}{4}a^{2}
\end{equation}
the above inequality yields the following quadratic equation
\begin{equation}\label{second degré 1}
	a^{2}+\left(2-\frac{4}{\nu}\right)a+\frac{4}{\nu}\geq 0
\end{equation}
Taking into account that $ \nu $ is a small parameter, the solution of Eq. \eqref{second degré 1} are $ a\lessapprox  \frac{1}{2}$ and $ a\gtrapprox \frac{4}{\nu}-\frac{5}{2} $. the first part of the solution is not acceptable since it corresponds to negative times, while the latest, using Eq. \eqref{amplitude at} and neglecting terms of order $ \nu^{2} $ , gives the breaking time $ t_{d,alv} $, as follows

\begin{equation}\label{t_d alternative}
	t_{d,alv}\approx t_b\left(1-\frac{\nu}{4}\right).
\end{equation}
Using Eq. \eqref{kdvb solition with dimensions},  the maximum height reached at $ t=t_{d,alv} $ is given by
\begin{equation}\label{a_max alternative}
	a_{max}(t_{d,alv})\approx 4h
\end{equation}
The wave length at $ t=	t_{d,a} $ is obtained from Eq. \eqref{lambda de t}, and reads
\begin{equation}
	\lambda(t=	t_{d,alv})=\left(\frac{1}{3}\right)^{1/2}h.
\end{equation}

\section{Integrability perspectives}

Recently, the integrability of a general form of the KdV and NLS equations with forcing involving product of eigenfunctions has been proved \cite{ocnmp:9884, Fokas_Latifi_2023}.

More specifically,  in \cite{ocnmp:9884} it is proved  that the following forced version of NLS is integrable:
\begin{align}\label{full NLS}
	\mathrm{i}q_t+\frac{\alpha}{2}q_{xx}-\alpha\lambda|q|^2q=\frac{2\mathrm{i}}{\pi}\int_{\mathbb{R}}\frac{g(t,k)}{a_0(k)}\Phi_1^+(x,t,k)&\Psi^+_1(x,t,k)\mathrm{e}^{\mathrm{i}\lambda \left(HG(t,k)|b_0|^2(k)\right)}dk,\nonumber\\
	& x\in\mathbb{R},\qquad t>0,
\end{align}
where $\alpha$ is a constant real parameter, $\lambda=\pm1$, $\Phi^+_1$, $\Psi^+_1$ are appropriate eigenfunctions,

\begin{align}\label{mathematical boundary conditions}
	G(t,k)=\int_0^tg(\tau,k)d\tau,\quad &b_0(k)=\lim_{x\rightarrow -\infty}\mathrm{e}^{2\mathrm{i}kx}\Psi^+_{10}(x,k),
	\nonumber \\
	&a_0(k)=\lim_{x\rightarrow -\infty}\Psi^+_{10}(x,k),
\end{align}

and $(\Psi_{10}^+(x,k), \Psi_{20}^+(x,k))$ are defined in terms
of $q_0(x)=q(x,0)$ by
\begin{align}\label{voltera}
	&\Psi_{10}^+(x,k)= -\int_x^\infty  d\xi \mathrm{e}^{2\mathrm{i}k(\xi-x)}q_0(\xi) \Psi_{20}^+(\xi,k), \nonumber\\ 
	&\Psi_{20}^+(x,k)=1-\int_x^\infty d\xi \lambda\overline{ q}_0(\xi) \Psi_{10}^+(\xi,k),
	\qquad -\infty<x<\infty \quad \text {Im}k\ge 0.
\end{align}
Eq. \eqref{full NLS} possesses the following Lax pair:
\begin{equation}\label{x evolution of X}
	X_x+\mathrm{i}k[\sigma_3,X]-QX=0,\qquad
	Q=\begin{pmatrix} 0 &  q \cr \lambda\overline{q} & 0 \end{pmatrix},
\end{equation}
\begin{align}\label{t evolution of X}
	X_t+&\mathrm{i}\alpha k^2[\sigma_3,X]-\alpha\left(kQ-\frac{\mathrm{i}}{2}(Q_x+Q^2)\sigma_3\right)X
	\nonumber\\
	&=\frac{1}{2\mathrm{i}}(Hg)X\sigma_3-\frac{1}{2\mathrm{i}}(HgX\sigma_3X^{-1})X,
	\qquad x\in\mathbbm{R}, \quad t>0, \quad k\in\mathbbm{R},
\end{align}

where $ Hf $  denotes the Hilbert transform
\begin{equation}
	(Hf)(k)=\frac{1}{\pi}p\int_{\mathbb{ R}}\frac{f(l)}{l-k}dl, \qquad k\in \mathbb{R}.
\end{equation}

And in \cite{Fokas_Latifi_2023}, it is shown that the forced integrable extension of the KdV, namely the following equation is integrable.

\begin{equation}\label{after differentiation Intro}
	u_t+\alpha(u_{xxx}+6uu_x)=d_x(x,t)+2h_x(x,t),
\end{equation}

where $ \alpha $ is introduced in order to consider the $ \alpha=0 $ limit and $ d $ and $ h $ are defined by
\begin{subequations}\label{definition of h and d Intro}
	\begin{align}
		&d(x,t)=\frac{1}{\pi}\int_\mathbb{R}g(t,l)(v_{11}v_{22}+v_{12}v_{21})(x,t,l)dl, \label{definition of d intro}\\
		&h(x,t)=\frac{1}{\pi}\int_\mathbb{R}g(t,l)v_{21}v_{22}(x,t,l)dl,\label{definition of h intro}
	\end{align}
\end{subequations}
is integrable. The functions $ v_{ij} $, $ i,j=1,2 $ are the $ ij $ components of the matrix $ v $ given by
\begin{equation}\label{v definition Intro}
	v(x,t,k)=\frac{1}{2}
	\begin{pmatrix} \phi(x,t,-k)-\displaystyle{\frac{1}{\mathrm{i}k}\phi_x(x,t,-k)} &  \phi(x,t,k)-\displaystyle{\frac{1}{\mathrm{i}k}\phi_x(x,t,k)} \cr
		\cr \phi(x,t,-k)+\displaystyle{\frac{1}{\mathrm{i}k}\phi_x(x,t,-k) } & \phi(x,t,k)+\displaystyle{\frac{1}{\mathrm{i}k}\phi_x(x,t,k)} \end{pmatrix},
\end{equation}
where $ \phi $ is an appropriate solution of the associated Lax pair. Namely, $ \phi $ satisfies 
\begin{equation}\label{x-part}
	\phi_{xx}+(u+k^2)\phi=0, 
\end{equation}
and
\begin{align}\label{again a monster Intro}
	&\phi_t+\alpha(4\mathrm{i}k^3-u_x)\phi+\alpha(2u-4k^2)\phi_x+\frac{1}{2\mathrm{i}}(Hg)\phi-\frac{1}{4\mathrm{i}}[Hg(\phi\widehat{\phi}+\frac{1}{k^2}\phi_x\widehat{\phi}_x)]\frac{\phi_x}{\mathrm{i}k}\nonumber\\
	&+\frac{1}{4\mathrm{i}}[Hg(\phi\widehat{\phi}-\frac{1}{k^2}\phi_x\widehat{\phi}_x)]\phi-\frac{1}{4\mathrm{i}}[Hg(\frac{1}{\mathrm{i}k}{(\phi\widehat{\phi})}_x)]\frac{\phi_x}{\mathrm{i}k}=0,
\end{align}
where the hat denotes  evaluation at $ -k $.

We have strong  indications to think that both the KdV-Burger equation \eqref{kdvb} and the NLS equation  \eqref{NLSforPsi}, with appropriate air and water dynamics and wind/wave interaction can be integrable. This will be the subject of fore coming studies.

%%%%%%%%%%%%%%%%%%%%%%%%%%%%%%%%%%%%%%%%%%%%%%%%%%%%%%%%%%%%%%%%%%%%%%%%%%%%%%%%%%%%%
%%%%%%%%%%%%%%%%%%%%%%%%%%%%%%%%%%%%%%%%%%%%%%%%%%%%%%%%%%%%%%%%%%%%%%%%%%%

\section{Conclusions}\label{sec4}

In the first part of this review paper, our aim was exclusively to derive a linear Miles' theory for waves propagating at finite depth $h$.  
Hence, we extended the well-known Miles' theory to the finite depth context under breeze to moderate wind conditions.
We have linearized the equations of motion governing the dynamics of the air/water interface problem
in finite depth and have studied the linear instability in time of a normal Fourier mode $k$. The prediction of exponential growth of wave amplitude (or energy) is well confirmed by field and laboratory experiments. As an extension, we have derived an anti-diffusive nonlinear Schr\"odinger equation in finite depth and found the Akhmediev,  Peregrine and Ma solutions for weak wind inputs in finite depth.

In the second part, We derived a SGN fully nonlinear, dispersive and focusing system of equations in the context of the nonlinear dynamics of surface water waves under wind forcing, in finite depth. We found that its 
soliton-like solution with amplitude, velocity, and effective wavelength increased with time. 
% We show that the theoretical growth in amplitude
% and the time of blow-up breaking are non-testable in an existing experimental facility.
Antidissipation due to wind action through the sheltering Jeffreys' mechanism
increases the amplitude of the solitary wave and leads to blow-up 
which occurs in finite time
for infinitely large asymptotic space.
The blow-up time is calculated. As an extension, the anti-diffusive Korteweg-de Vries-Burger equation is derived.  we have studied in detail the kinetics of the breaking of the wind/wave solitary wave and reviewed various break-up criteria.
Experimentally, it is clear that the breaking will occur before the blow up.   The experimental confirmation of the present theory can be tested in existing  facilities. We will conduct further investigations in this direction in forecoming studies.

Finally, the perspectives of integrability for the NLS equation and the KdV-B equation have been exposed which also be investigated  in forecoming studies.

\appendix

\section*{Appendix: Direct derivation of the anti-diffusive Korteweg-de Vries-Burger equation from Euler Equations}\label{A}

\setcounter{section}{1}
\newtheorem{Proposition}{Proposition}[section]

Let us consider
{\it a quasi-linear} air/water system with the air dynamics linearized and the water dynamics considered nonlinear and irrotational.
The system is $(2+1)$ dimensional ($x,z,t$) with $x$ and $z$ the vertical and the horizontal space coordinates.
The aerodynamic air pressure $P_a(x,z,t)$ evaluated at the free surface $z=\eta(x,t)$ 
has a component in phase and a component in quadrature with the water elevation. For an
energy flux to occur from the wind to the water waves there must be
a phase shift between the fluctuating pressure and the interface.
Hence, the energy transfer is only due to the component in
quadrature with the water surface, or in other words in phase with
the slope. To simplify the problem we consider, following references \cite{Jeffreys2, Miles1, Kraenkel}, only the pressure
component in phase with the slope on the interface i.e.,
\begin{equation}
	P_a=\epsilon \rho_{a}\Delta^2\eta_{x}\quad \mbox{with}\quad \Delta=[\frac{\kappa U_{1}}{\sqrt{C_{10}}}-c] 
\end{equation}
where $\epsilon<1$ is the sheltering coefficient, $c=\sqrt{g/k}\tanh(kh)$, $U_1=u_{*}/\kappa$
with $u_{*}$ the friction velocity, $\kappa \sim 0,41$ the Von K\`arm\`an constant, $C_{10}$ the wind-stress coefficient and $g$ the gravitational
acceleration. This is nothing more that Jeffrey's sheltering mechanisms.
In order to adimensionalize the equations of motion
we introduce dimensionless primed variables: 
$x=lx', z=hz', t=lt'/c_0, \eta=a\eta', \phi=gla\phi'/c_0,
U_1=c_0 U_1'$
with $\phi$ the velocity potential and $a$ and $l$ typical wave amplitude and wavelength and 
$c_{0}=\sqrt{gh}$. We define two dimensionless parameters
$\nu=a/h < 1$ and $\delta=h/l<1.$
So with this assumption
the complete irrotational Euler equations and boundary conditions
are (dropping the primes)
\begin{eqnarray}
	&&\delta^2\phi_{xx} + \phi_{zz}=0, \quad-1 \le z \le \nu\eta,\label{adlaplace}\\
	&&\phi_z =0,\quad z=-1, \label{adBClaplace}\\
	&&\eta_t + \nu\phi_x\eta_x -\frac{1}{\delta^2}\phi_z = 0, \quad z= \nu\eta, \label{adBCkine} \\
	&&\phi_t + \frac{\nu}{2}\phi_x^2 + \frac{\nu}{2\delta^2}\phi_z^2 +
	\eta + \delta \epsilon s \Delta^2\eta_x=0,  \quad z=\nu \eta,
	\label{adBCdyn}
\end{eqnarray}
where $ s=\rho{_a}/\rho_{w}\sim 10^{-3}$ with $\rho_a$ ($\rho_w$) the air (water) density. 
We solve the Laplace equation and its boundary conditions
with an expansion in powers of $(z + 1)$, namely
\begin{equation}\label{expansion}
	\phi=\sum_{m=0}^{m=\infty}(z+1)^m\delta^mq_{m}(x,t).
\end{equation}

Substituting \eqref{expansion} in Eq.(\ref{adlaplace}) and using Eq.(\ref{adBClaplace}) we obtain 
\begin{equation}
	\phi=\sum_{m=0}^{m=\infty}(-1)^m\frac{(z+1)^{2m}}{(2m)!}\delta^{2m}
	q_{0,2mx}.
\end{equation}
Using the kinematic and dynamics boundary conditions Eq.(\ref{adBCkine})
and Eq.(\ref{adBCdyn}) and
disregarding terms in $\mathcal{O}(\nu\delta^{2})$ and $\mathcal{O}(\delta^{4})$ we find, with $r=q_{0,x}$,
the system
\begin{eqnarray}
	&&\eta_t+\{(1+\nu\eta)r\}_x-\frac{1}{6}\delta^2 r_{xxx} = 0, \label{mboussi1}\\
	&&\eta_x+r_t+\nu rr_x-\frac12  \delta^2r_{xxt}
	+\delta \epsilon s \Delta^2\eta_{xx}= 0.\label{mboussi2}
\end{eqnarray}
The linear wave solution of (\ref{mboussi1}) and (\ref{mboussi2}) moving to the right is
$r(\xi) =\eta(\xi)$, $ \xi=x-t$,
with $\eta$ (or $r$) an arbitrary function of $\xi$.
Now we look for a solution with nonlinear corrections to the orders $\mathcal{O}(\nu)$, $\mathcal{O}(s\delta)$, and
$\mathcal{O}(\delta^2)$. Following procedure in reference \cite{Whitham} we obtain
\begin{eqnarray}
	\label{functionr}
	r=\eta-\frac{1}{4}\eta^2 \nu+\frac{\epsilon}{2}\Delta^2\eta_{x} s\delta+\frac{1}{3}\eta_{xx} \delta^2  + \mathcal{O}(\nu\delta^2,s^2\delta^2,\delta^4),
\end{eqnarray}
Substituting (\ref{functionr}) in (\ref{mboussi1}) and (\ref{mboussi2}) we obtain
a focusing KdV-B equation
\begin{equation}\label{dimensionless}
	\eta_t+\eta_x+\frac32\nu\eta\eta_x+\frac16 \delta^2\eta_{xxx}+\frac{s}{2}\delta \epsilon  \Delta^2\eta_{xx}=0.
\end{equation}
For traveling wave solutions, the action of dissipation or "anti-dissipation" in KdV-B is not of great matter except for the sign of the slope \cite{JeffreyKdVB}.
But the soliton-like solutions under the continuous energy transfer from wind to water, exhibits a blow-up and breaking in finite time.
%%%%%%%%%%%%%%%%%%%%%%%%%%%%%%%%%%%%%%%%%%%%%%%%%%%%%%%%%%%%%%%%%%%%%%%%%%%%%%%%%%%%%%%%%%%%%

\label{lastpage}

\begin{thebibliography}{99}


		
\bibitem{Janssen} Janssen, P. The Interaction of Ocean Waves and Wind (Cambridge University Press, \textbf{2004})

\bibitem{Jeffreys1}
Jeffreys, H.
\newblock On the formation of water waves by wind,
\newblock {\em Proc. R. Soc.} { A107}, {\bf 1925}

\bibitem{Jeffreys2}
Jeffreys, H.
\newblock On the formation of water waves by wind (Second paper),
\newblock {\em Proc. R. Soc.} {A110}, {\bf 1926}



\bibitem{Phillips}
Phillips, O. 
\newblock  On the generation of waves by turbulent wind, {\em J. Fluid Mech.} {2},  \textbf{1957}


\bibitem{Miles2}
Miles, J. Generation of surface waves by winds, {\em Appl. Mech. Rev}. {50},  \textbf{1997}

\bibitem{Janssen2}Janssen, P. Quasi-linear theory of wind-wave generation applied to wave forecast, {\em J. Phys. Oceanogr.} {21}, \textbf{1991}



\bibitem{Belcher}Belcher, S. \& Hunt, J. Turbulent shear flow over slowly moving waves, {\em J. Fluid Mech.} {251}, \textbf{1993}





\bibitem{fluids7080266}
Manna, M. \& Latifi, A. ~Serre-Green-Naghdi Dynamics under the Action of the Jeffreys' Wind-Wave Interaction, {\em Fluids}, {7}, \textbf{2022}, https://www.mdpi.com/2311-5521/7/8/266


\bibitem{JAFD_2017}
Latifi, A., Manna, M.A., Montalvo, P. \& Ruivo, M.
\newblock Linear and Weakly Nonlinear Models of Wind Generated Surface Waves in
Finite Depth, 
\newblock {\em J. Appl. Fluid Mech.} { 10}, {\bf 2017}, 
\newblock {{https://doi.org/10.29252/jafm.73.245.27597}}

\bibitem{Manna_2018}
Manna, M.A., Latifi, A. \& Kraenkel, R.A.
\newblock Green--Naghdi dynamics of surface wind waves in finite
depth, 
\newblock {\em Fluid Dyn. Res.}  {50}, {\bf 2018},
\newblock {{https://doi.org/10.1088/1873-7005/aaa739}}



\bibitem{fluids8080231}Manna, M. \& Latifi, A. Korteweg–De Vries–Burger Equation with Jeffreys’ Wind–Wave Interaction: Blow-Up and Breaking of Soliton-like Solutions in Finite Time, {\em Fluids}, {8}, \textbf{2023}, https://www.mdpi.com/2311-5521/8/8/231



\bibitem{YoungVerhagen1}Young, I. The growth rate of finite depth wind-generated waves, {\em Coastal Engineering}. {32},  \textbf{1997}

\bibitem{YoungVerhagen2}Young, I. \& Verhagen, L. The growth of fetch limited waves in water of finite depth, Part 2: Spectral evolution, {\em Coastal Engineering}, {29}, \textbf{1996}


\bibitem{IjimaTang}Ijima, T. \& Tang, F. Numerical calculation of wind waves in shallow water, {\em Coastal Engineering Proceedings, North America}, \textbf{2011}, https://journals.tdl.org/ICCE/article/view/2401/2076

\bibitem{Lighthill1978-fo}Lighthill, J. Waves in Fluids (Cambridge University Press, \textbf{1978})







\bibitem{Charnock}Charnock, H. Wind stress on a water surface, {\em Quart. J. Roy. Meteorol. Soc.} {81},  \textbf{1955}

\bibitem{Fairall}Fairall, C., Grachev, A., Bedard, A. \& Nishiyama, R. Wind, wave, stress, and surface roughness relationships from turbulence measurements made on R/P flip in the scope experiment, {\em NOAA Technical Memorandum}, {ERL ETL-268},  \textbf{1996}


\bibitem{Garratt}Garratt, J., Hess, G., Physick, W. \& Bougeault, P. The atmospheric boundary layer advances in knowledge and application, {\em Boundary-layer Meteorology},  {78}, \textbf{1996}

\bibitem{Tennekes}Tennekes, H. The logarithmic wind profile, {\em J. Atmos. Sc.} {30}, \textbf{1972}

\bibitem{Rayleigh}Rayleigh, L. On the stability or instability of certain fluid motions, {\em Proc. Lond. Math. Soc.} {XI}, \textbf{1880}


\bibitem{Thijsse}Thijsse, J. Dimensions of wind-generated waves, {\em General Assembly Of Association D'Océanographie Physique, Procés-Verbaux}, {4}, \textbf{1949}


\bibitem{Bretschneider}Bretschneider, C. Revised wave forecasting relationships, {\em  Proc. 6th Conference On Coastal Engineering, Gainsville/Palm Beach/Miami Beach, FL.} ( {ASCE, New York, \textbf{1958}} ) 


\bibitem{Young1}
Young, I. The growth rate of finite depth wind-generated waves, {\em Coastal Eng.} {32},  \textbf{1997}

\bibitem{Young1999-zm}Young, I. Wind generated ocean waves (Elsevier Science \& Technology, \textbf{1999})





\bibitem{Donelan}
Donelan, M., Babanin, A., Young, I. \& Banner, M. Wave-Follower field measurements of the wind-input spectral function, 
Part II: Parameterization of the wind input, 
{\em Journal Of Physical Oceanography}, {36}, \textbf{2006}


\bibitem{PiersonMo}Pierson, W. \& Moskowitz, L. A proposed spectral form for fully developed wind seas based on the similarity theory of S.A. Kitaigorodskii, {\em Journal Of Geophysical Research}, {69},  \textbf{1964}


\bibitem{JinWu}Wu, J. Wind-Stress Coefficients over sea surface from breeze to hurricane, {\em J. Geophy. Res.} {87}, \textbf{1982}


\bibitem{Fenton}Fenton, J. A high-order cnoidal wave theory, {\em J. Fluid Mech.} {94}, \textbf{1979}


\bibitem{Marcus}Francius, M. \& Kharif, C. Three-dimensional instabilities of periodic gravity waves in shallow water,  {\em J. Fluid Mech.} {561}, \textbf{2006}

\bibitem{Miles1}Miles, J. On the generation of surface waves by shear flows, {\em J. Fluid Mech}. {3}, \textbf{1957}


\bibitem{Thomas}
Thomas, R., Kharif, C. \& Manna, M.A. nonlinear Schrödinger equation for waves on finite depth with constant vorticity,
{\em Phys. Fluids}, \textbf{2012}

\bibitem{Touboul1}Touboul, J., Kharif, C., Pelinovsky, E. \& Giovanangeli, J. On the interaction of wind and steep gravity wave groups using Miles' and Jeffreys' mechanisms, {\em Nonlin. Processes Geophys.} {15}, \textbf{2008}

\bibitem{Touboul2}Touboul, J., Giovanangeli, J., Kharif, C. \& Pelinovsky, E. Freak waves under the action of wind: experiments and simulations, {\em Eur. J.  Mech. B: Fluids}, {25}, \textbf{2006}

\bibitem{Kharif}Kharif, C., Giovanangeli, J., Touboul., C., Grade, L. \& Pelinovsky, E. Influence of wind on extreme wave events: experimental and numerical approaches, {\em J. Fluid Mech.} {594}, \textbf{2008}

\bibitem{Onorato}Onorato, M. \& Proment, D. Approximate rogue wave solutions of the forced and damped Nonlinear Schrödinger Equation fro water waves, {\em Phy. Lett. A}, 376, \textbf{2012}

\bibitem{Akhmediev}Akhmediev, N., Eleonskii, V. \& Kulagin, N. Exact first-order solution on the nonlinear Schrödinger equation, {\em Theor. Math. Phys.} {72}, \textbf{1987}

\bibitem{Peregrine}Peregrine, D. Water waves, nonlinear Schrödinger equation and their solutions {\em J. Austral. Math. Soc. Ser. B}, {25}, \textbf{1983}

\bibitem{Ma}Ma, Y. The perturbed plane-wave solutions of the cubic Schrödinger equation, {\em Stud. Appl. Math.} {60}, \textbf{1979}

\bibitem{Dysthe}Dysthe, K. \& Trulsen, K. Note on Breather Type Solutions of the NLS as Models for Freak-Waves, {\em Physica Scripta}, {82}, \textbf{1999}

\bibitem{Grinevich_2021}
Grinevich, P. \& Santini, P.  The linear and nonlinear instability of the Akhmediev breather, {\em Nonlinearity}, {34}, \textbf{2021}, https://dx.doi.org/10.1088/1361-6544/ac3143


\bibitem{10.1063/1.870058}Reul, N., Branger, H. \& Giovanangeli, J. Air flow separation over unsteady breaking waves, {\em Physics Of Fluids}, {11}, \textbf{1999}, https://doi.org/10.1063/1.870058

\bibitem{Su}Su, C. \& Gardner, C. Collisional theory of shock and nonlinear waves in plasma, {\em J. Math. Phys.} {10}, \textbf{1969}


\bibitem{Serre}
Serre, F.
\newblock Contribution à L’étude des écoulements Permanents et Variables
Dans Les Canaux,
\newblock {\em La Houille Blanche}, 3, \textbf{1953}

\bibitem{bams/1183416937}
Dunkel, O.~{Some applications of Green's theorem in one dimension},~{\em Bull. Am. Math. Soc.} 8, \textbf{1902}, https://doi.org/bams/1183416937

\bibitem{doi:10.1142/5740}
Svendsen, I.A.
\newblock {\em Introduction to Nearshore Hydrodynamics} (World Scientific: {London, UK,} 
\textbf{2005}), 
\newblock {{https://doi.org/10.1142/5740}}

\bibitem{doi:10.1142/0752}
Chiang, C.M.
\newblock {\em The Applied Dynamics of Ocean Surface Waves} (World Scientific: {London, UK,}\textbf{ 1992}),
\newblock{{https://doi.org/10.1142/0752}}

\bibitem{DarbouxLeonsSL}Darboux, G. \textit{Leçons sur la Théorie Générale des Surfaces et les Applications Géométriques Du Calcul Infinitésimal} (Legare Street Press, \textbf{2023})


\bibitem{doi:10.1063/1.1692358}
Ott, E.; Sudan, R.N.
\newblock Nonlinear Theory of Ion Acoustic Waves with Landau Damping,
\newblock{\em Phys. Fluids}, 12, {\bf 1969}, 
\newblock{{https://doi.org/10.1063/1.1692358}}.



\bibitem{doi:10.1063/1.1693097}
Ott, E. \& Sudan, R.N.
\newblock Damping of Solitary Waves,
\newblock {\em Phys. Fluids}, 13, {\bf 1970}, 
\newblock {{https://doi.org/10.1063/1.1693097}}


\bibitem{GN2}
Green, A. \& Naghdi, P.M.~A Derivation of Equations for Wave Propagation in Water of Variable Depth, ~{\em Fluid. Mech.} 78, {\bf 1976}


\bibitem{BRANGER2022104174}
Branger, H., Manna, M., Luneau, C., Abid, M. \& Kharif, C.
\newblock Growth of surface wind-waves in water of finite depth: A laboratory
experiment,
\newblock {\em Coast. Eng.} 177, {\bf 2022}, 
\newblock
{{https://doi.org/https://doi.org/10.1016/j.coastaleng.2022.104174}}


\bibitem{Whitham}
Whitham, G.
\newblock {\em Linear and Nonlinear Waves}  ({Wiley: New York, NY, USA,} 
\textbf{1974})


\bibitem{KdV}
Korteweg, D. \& de~Vries, G.
\newblock On the Change of Form of Long Waves Advancing in a Rectangular Canal,
and on a New Type of Long Stationary Waves,
\newblock {\em Philos. Mag.} 39, {\bf 1895}

\bibitem{Benney}
Benney, D.J.
\newblock Long waves in liquid films,
\newblock {\em J. Math. Phys.} 45, {\bf 1996}

\bibitem{Johnson}
Johnson, R.
\newblock Shallow water waves on a viscous fluid-the undular bore,
\newblock {\em Phys. Fluids}, 15,  {\bf 1972}

\bibitem{Grad1}
Grad, H. \& Hu, P.
\newblock Unified shock in plasma,
\newblock {\em Phys. Fluids}, 10, {\bf 1967}

\bibitem{Grad2}
Hu, P.
\newblock Collisional theory of shock and nonlinear waves in plasma,
\newblock {\em Phys. Fluids}, 15, {\bf 1972}

\bibitem{Wadati}
Wadati, M.
\newblock Wave propagation in nonlinear lattice,
\newblock {\em J. Phys. Soc. Jpn.} 38, {\bf 1975}

\bibitem{Kara}
Karahara, T.
\newblock Weak nonlinear magneto-acoustic waves in a cold plasma in the
presence of effective electron-ion collisions,
\newblock {\em J. Phys. Soc. Jpn.} 27, {\bf 1970}






\bibitem{GN1}
Green, A., Laws, N. \& Naghdi, P.M.
\newblock On the theory of water waves,
\newblock {\em Proc. R. Soc. A}, 338, {\bf 1974}

\bibitem{McCowan}
McCowan, J.
\newblock On the highest wave of permanent type,
\newblock {\em Philos. Mag. Ser. 5}, 38, {\bf 1894}

\bibitem{Miche}
Miche, R.
\newblock {\em Mouvement Ondulatoires de la Mer en Profondeur Constante ou Décroissante} ({École Nationale des Ponts et Chaussées}: {Marne-la-Vallée, France}, \textbf{1944})

\bibitem{Shemer}
Shemer, L.
\newblock On kinematics of very steep waves,
\newblock {\em Nat. Hazards Earth Syst. Sci.} 13, {\bf 2013}




\bibitem{ocnmp:9884}Fokas, A. \& Latifi, A. The nonlinear Schrödinger equation with forcing involving products of eigenfunctions, {\em Open Comm. Nonlinear Math. Phy.} 2, \textbf{2022}, https://ocnmp.episciences.org/9884

\bibitem{Fokas_Latifi_2023}Fokas, A. \& Latifi, A. The Korteweg–De Vries Equation with Forcing Involving Products of Eigenfunctions, {\em J. Of Math. Phy., Anal., Geom.} {19},  \textbf{2023}, https://jmag.ilt.kharkiv.ua/index.php/jmag/article/view/998


\bibitem{Kraenkel}Kharif, C., Kraenkel, R., Manna, M. \& Thomas, R. The modulational instability in deep water under the action of wind and dissipation, {\em J. Fluid Mech.} {664}, \textbf{2010}

\bibitem{JeffreyKdVB}Jeffrey, A. \& Xu, S. Exact solutions of the Korteweg-de Vries-Burgers equation, {\em Wave Motion}, {11}, \textbf{1989}







\end{thebibliography}
\end{document}